\documentclass{article}

\usepackage{arxiv}

\usepackage[utf8]{inputenc} 
\usepackage[T1]{fontenc}    
\usepackage{hyperref}       
\usepackage{url}            
\usepackage{booktabs}       
\usepackage{amsfonts}       
\usepackage{nicefrac}       
\usepackage{microtype}      
\usepackage{lipsum}
\usepackage{graphicx}
\usepackage{amsmath}
\usepackage{amsfonts}
\usepackage{natbib}
\def\bSig\mathbf{\Sigma}

\title{An adaptive enrichment design using Bayesian model averaging for selection and threshold-identification of tailoring variables}

\author{
 Lara Maleyeff \\
  Department of Epidemiology, Biostatistics, and Occupational Health, McGill University, Montr\'eal, QC, CA \\
  \texttt{lara.maleyeff@mcgill.ca} \\
   \And
 Shirin Golchi \\
  Department of Epidemiology, Biostatistics, and Occupational Health, McGill University, Montr\'eal, QC, CA \\
  \And
 Erica E.M. Moodie\\
  Department of Epidemiology, Biostatistics, and Occupational Health, McGill University, Montr\'eal, QC, CA \\ 
  \And
  Marie Hudson \\
  Department of Medicine, McGill University, Montr\'eal, QC, CA \\ 
  Jewish General Hospital and Lady Davis Institute for Medical Research, Montr\'eal, QC, CA
}

\begin{document}

\label{firstpage}

\maketitle

\begin{abstract}
Precision medicine stands as a transformative approach in healthcare, offering tailored treatments that can enhance patient outcomes and reduce healthcare costs. As understanding of complex disease improves, clinical trials are being designed to detect subgroups of patients with enhanced treatment effects. Biomarker-driven adaptive enrichment designs, which enroll a general population initially and later restrict accrual to treatment-sensitive patients, are gaining popularity. Current practice often assumes either pre-trial knowledge of biomarkers defining treatment-sensitive subpopulations or a simple, linear relationship between continuous markers and treatment effectiveness. Motivated by a trial studying rheumatoid arthritis treatment, we propose a Bayesian adaptive enrichment design which identifies important tailoring variables out of a larger set of candidate biomarkers. Our proposed design is equipped with a flexible modelling framework where the effects of continuous biomarkers are introduced using free knot B-splines. The parameters of interest are then estimated by marginalizing over the space of all possible variable combinations using Bayesian model averaging. At interim analyses, we assess whether a biomarker-defined subgroup has enhanced or reduced treatment effects, allowing for early termination due to efficacy or futility and restricting future enrollment to treatment-sensitive patients. We consider pre-categorized and continuous biomarkers, the latter of which may have complex, nonlinear relationships to the outcome and treatment effect. Using simulations, we derive the operating characteristics of our design and compare its performance to two existing approaches.
\end{abstract}

%


\maketitle


%

\section{Introduction}
\label{s:intro}

Precision medicine, which allows for decisions about medical interventions to depend on patient characteristics, may be more efficient, less costly, and result in better patient outcomes, on average, than a traditional one-size-fits-all approach.  This is particularly true in domains where response to treatment is highly heterogeneous, such as in rheumatoid arthritis (RA) and oncology. For example, of the several treatment options available for RA, 30\% of patients  fail any particular drug \citep{smolen2018}. The current trial-and-error approach in RA, in which patients attempt multiple drugs over the span of months or years until their disease is controlled, results in unnecessary comorbidity and potential patient harm. In the presence of such heterogeneous treatment effects, traditional clinical trials, which detect a population-averaged effect, may be inefficient \citep{adams2006estimating}. As molecular phenotyping becomes increasingly available and inexpensive to obtain, clinical trials to assess treatment efficacy in biomarker-defined subgroups are becoming more feasible. Such validated, quality information is crucial to determine appropriate personalized treatments. In addition, incorporating  statistically sound subgroup-identification methods can lessen the risk of conducting null trials and reduce exposure of patients to treatments which do not benefit them. 


In the clinical trial setting, several biomarker-driven designs exist to detect heterogeneous effects. Adaptive enrichment designs initially enroll and randomize patients regardless of their biomarker status, then use interim data to cease enrollment of certain biomarker-defined subgroups for efficacy or futility. \cite{wang2007approaches} developed a two-stage adaptive design to assess treatment effects across levels of a dichotomous biomarker; multiple similar extensions have been proposed \citep{karuri2012two, magnusson2013group,rosenblum2016group, brannath2009confirmatory, jenkins2011adaptive,mehta2014biomarker}. Basket trials have become more prolific in recent years due their ability to cost-effectively evaluate the activity of a single treatment on groups of patients with varying biomarker status, or ``baskets" \citep{asano2020bayesian,chen2020bayesian, chu2018bayesian, jin2020bayesian, park2020overview}. In such trials, the goal is to determine the subset of baskets in which treatment is most effective. Development of these designs is a rich area of active research; a comprehensive review can be found in \citep{simon2017critical}. \cite{psioda2021bayesian} developed a Bayesian basket trial design methodology which uses Bayesian model averaging (BMA). BMA, a coherent mechanism for accounting for inherent uncertainly in model selection, yields posterior distributions which are marginalized over a space of submodels. This approach naturally allows for information borrowing by weighing multiple models based on their posterior probabilities, leading to more robust and informed decision-making.

Most biomarker-driven designs, including those described above, focus on pre-determined patient subgroups.  Such subgroups are defined by biomarkers which are categorized, either as levels of naturally categorized variables or by using pre-specified thresholds. However, there exist some trial designs which incorporate biomarker identification. \cite{freidlin2005adaptive} and \cite{freidlin2010cross} propose a two-stage approach, wherein the first stage machine learning techniques are used to identify a treatment-sensitive population; in the second stage, the effectiveness of treatment is assessed for the sensitive group. There also exist designs which incorporate variable selection at the interim analysis stage. \cite{gu2016bayesian} propose a multi-stage design, called BATTLE-2, in which Bayesian lasso is used at interim analyses to refine the set of biomarkers considered in the remainder of the trial. \cite{park2022bayesian} propose a Bayesian group sequential enrichment design which incorporates variable selection to reduce the space of candidate biomarkers and uses an early response proxy outcome that is used in interim analyses. At each interim decision, spike-and-slab priors are used to identify a set of tailoring biomarkers that characterizes treatment-sensitive patients and decisions are made to stop the trial for efficacy or futility. Subsequent enrollment is limited to the most recently identified treatment-sensitive subgroup. Their methods can accommodate both categorical and continuous biomarkers, assuming continuous biomarkers have a monotone, linear relationship with treatment effect. This assumption is commonly made in much of the threshold identification literature \citep{jiang2007biomarker, renfro2014adaptive, ohwada2016bayesian}. 

Recently, \cite{liu2022bayesian} developed a Bayesian adaptive enrichment design which allows for a nonlinear and nonmonotone relationship between one or more continuous biomarkers and treatment. Rather than utilizing algorithmic dichotomization to determine biomarker thresholds, they flexibly estimate continuous biomarker relationships directly and quantify the posterior uncertainty around these estimates. They specify an interim analysis procedure in which subgroups are identified and assessed for efficacy or futility. If no subgroup meets the pre-specified criteria, an overall analysis is performed. In simulation studies, their design had similar or improved operating characteristics compared with more traditional grid-search based approaches. While simulation studies showed positive performance, this method does not incorporate variable selection; that is, the set of tailoring biomarkers is fixed and assumed to be known \textit{a priori}.

We propose a Bayesian adaptive enrichment clinical trial which identifies important tailoring variables out of a larger set of candidate biomarkers, flexibly models the effects of continuous biomarkers using free knot B-splines, and marginalizes over the space of all possible variable combinations, reducing dependence on one single model specification. At interim analysis, we assess whether a biomarker-defined subgroup, which may be the entire trial population, has enhanced or reduced treatment effects, allowing for early stoppage due to efficacy or futility. At the end of the trial, the set of candidate tailoring variables is refined using posterior inclusion probabilities and a reduced model is used for treatment recommendations. The novelty of our work lies in the combination of flexibly modelling continuous biomarker-treatment effect relationships using free knot splines, while simultaneously using BMA to reduce the set of candidate biomarkers and marginalize over the space of all possible submodels. We compare our proposed method with two state-of-the-art comparators, \cite{park2022bayesian} and \cite{liu2022bayesian}. Both of these methods out-perform existing enrichment designs. Comparison with \cite{park2022bayesian} emphasizes the importance of accounting for nonlinear and nonmonotone relationships between biomarkers and treatment effectiveness and comparison with \cite{liu2022bayesian} highlights the effect of variable selection. 

The remainder of the paper is organized as follows. Section \ref{s:design} describes the notation, models, and interim analysis algorithm associated with the proposed design. We also describe the computational methods used to derive the posterior distribution of all unknown parameters. Section \ref{s:data_example} describes a hypothetical trial to detect treatment-sensitive RA patients, motivating the simulation study in Section \ref{sec:sims}, in which we assess the operating characteristics of the proposed design under a variety of settings and compare its performance to two existing methods. We finish with a discussion in Section \ref{s:discussion}.

\section{Methodology}
\label{s:design}

In this section, we propose analytic methods, appropriate for both Bayesian adaptive enrichment designs and broader applications, to determine a precision medicine strategy in the presence of multiple candidate biomarkers, which may be dichtomous or continuous. To this end, we first define a class of appropriate models, that flexibly relate the treatment, predictive and tailoring variables, and their interactions to the study outcome of interest. We then use BMA to derive a posterior distribution which accounts for the heterogeneity inherent in selecting an appropriate model. Using the derived posterior distribution, we can then identify the range of values where treatment has a positive, clinically important effect. 

\subsection{Notation}
For $i=1,\dots,n$, let $Y_i$ be the study outcome for individual $i$, $T_i$ be a treatment indicator for individual $i$ ($T_i=0$ if assigned to control, $T_i=1$ if assigned to treatment), $\mathbf{x}_i = (x_{1i},\dots,x_{Ji})^\top$ be a set of continuous predictive variables for $Y_i$, $\mathbf{z}_i=(z_{1i},\dots,z_{Ri})^\top$ be a set of candidate binary predictive variables for $Y_i$, and $\mathbf{\tilde{x}}_i$,
$\mathbf{\tilde{z}}_i$ be sets of candidate continuous and binary tailoring variables, respectively. Throughout, \textit{tailoring} variables refer to those which interact with treatment to modify treatment effects within specific subgroups, while a \textit{predictive} variable directly influences the outcome regardless of interactions. For both binary and continuous variables, the set of candidate tailoring variables is constrained to be a subset of the set of candidate predictive variables. We denote $\mathcal{D} = \{(Y_i, T_i, \mathbf{x}_i, \mathbf{z}_i)\text{, } i=1,\dots,n\}$ as the complete set of information available for all individuals in the study. 

\subsection{A fully Bayesian model specification}

In order to identify a subgroup of individuals where treatment is most effective, we first postulate a fully Bayesian model, 

\begin{equation}
\label{eq:complete}
    g(\mathbb{E}\left[Y_{i} \mid \mathbf{x}_i, \mathbf{z}_i, T_i\right]) = \mu + \sum_{j=1}^{J} h_{1,j}(x_{ji}) + \boldsymbol{\beta}_1^T \mathbf{z}_i + \left(\phi + \sum_{j=1}^{\tilde{J}} h_{2,j}(\tilde{x}_{ji}) + \boldsymbol{\beta}_2^T \mathbf{\tilde{z}}_i\right) T_i,
\end{equation}
where $g$ is a pre-specified link function, $\mathbb{E}\left[Y_{i} \mid \mathbf{x}_i, \mathbf{z}_i, T_i\right]$ is the conditional mean of the outcome, $\mu$ is the intercept, $\phi$ is the main effect of treatment, $h_{1,s}(\cdot)$, $s=1,\dots,J$ and  $h_{2,s+J}(\cdot)$, $s=1,\dots,\tilde{J}$ are free knot splines with pre-specified candidate knots $K_s$ corresponding to the predictive and tailoring effects of the candidate continuous biomarkers, respectively, and $\boldsymbol{\beta}_1$ and $\boldsymbol{\beta}_2$ are the predictive and tailoring effects of the candidate binary biomarkers, respectively. Additionally, we denote the coefficients associated with spline $s$ as $\boldsymbol{\theta}_s$ and the complete set of model coefficients as $\boldsymbol{\zeta}=(\mu,\phi, \boldsymbol{\theta}_1^\top,\dots, \boldsymbol{\theta}_{J}^\top, \boldsymbol{\beta}_1^\top, \boldsymbol{\theta}_{J+1}^\top,\dots, \boldsymbol{\theta}_{J+\tilde{J}}^\top,\boldsymbol{\beta}_2^\top)^\top$. We note that the model formulation described in \eqref{eq:complete} contains all candidate tailoring variables, representing the postulated ``full model" in the proposed procedure. Throughout the paper, we refer to the proposed procedure as ``fitting Model \eqref{eq:complete}"; this refers to the computation of the posterior distribution averaged over the space of candidate submodels and sub-splines.

We differentiate between model \textit{coefficients}, \textit{terms}, and \textit{variables}. \textit{Variables} refer to the biomarkers included in the model. These variables may be predictive (included in the main effect), tailoring (included in the interaction effect), or both. Within a model, a variable may be included in up to two \textit{terms}: namely, a main effect term and a tailoring effect term. During the model fitting procedure, terms are iteratively added or removed from the model. For example, the interaction effect associated with age can be removed, while the main effect of age remains. Conversely, the main effect of age cannot be removed without first removing its interaction effect to maintain well-formulated model hierarchy. A term may consist of one or more \textit{coefficients}. While terms associated with binary variables contain a single coefficient, terms associated with continuous variables generally contain multiple coefficients that describe the spline function. For example, suppose we have continuous age and binary sex as candidate predictive and tailoring variables. Using a 3-degree B-spline with no knots and excluding the intercept and main effect of treatment, this model has 2 variables (age and sex), 4 associated terms (the main and interaction effect of each), and 8 coefficients (3 for each spline term, 1 for each binary term).

In order to detect heterogeneous treatment effects, a common goal is to identify the effective subspace. First, we let $\boldsymbol{\gamma}(\mathbf{\tilde{x}},\mathbf{\tilde{z}} \mid \phi, \boldsymbol{\theta}_{J+1},\dots,\boldsymbol{\theta}_{J+\tilde{J}}, \boldsymbol{\beta}_2) = \phi + \sum_{j=1}^{\tilde{J}} h_{2,j}(\tilde{x}_{ji}) + \boldsymbol{\beta}_2^T \mathbf{\tilde{z}}$ be the variable-specific treatment effect or the blip effect of treatment \citep{robins1997causal}; we omit the conditioning and use the shorthand $\boldsymbol{\gamma}(\mathbf{\tilde{x}},\mathbf{\tilde{z}})$ throughout. Then the effective subspace can be defined as
\begin{equation}
\label{eq:eff_subspace}
    \mathcal{X}^* = \left\{\mathbf{\tilde{x}}, \mathbf{\tilde{z}}: P\left( \boldsymbol{\gamma}(\mathbf{\tilde{x}},\mathbf{\tilde{z}}) > e_1 \mid \mathcal{D}\right)>1-\alpha\right\},
\end{equation}
where thresholds $\alpha$ and $e_1$ are predefined; $\alpha$ can be tuned through simulation, while $e_1$ is typically determined based on clinical expertise. Given the effective subspace, the hypothesis of a trial designed to detect a treatment effect in the effective subspace can be written as 
\begin{equation*}
    H_0: \gamma(\mathbf{\tilde{x}},\mathbf{\tilde{z}}) \le e_1 \text{ } \forall \text{ }\mathbf{\tilde{x}},\mathbf{\tilde{z}} \in \mathcal{X}  \text{ vs. } H_A: \exists \text{ } \mathbf{\tilde{x}},\mathbf{\tilde{z}} \in \mathcal{X}  \text{ s.t. } \gamma(\mathbf{\tilde{x}},\mathbf{\tilde{z}}) > e_1,
\end{equation*}
where $\mathcal{X}$ is the complete space of candidate tailoring variables. In other words, the null hypothesis is that the treatment is ineffective for all variable combinations and the alternative hypothesis is that there exists at least one variable combination for which the treatment is effective. 

We then assign a prior probability to each submodel. A submodel containing $m$ biomarker terms is assumed to follow a right-truncated Poisson distribution at $p$, where $p$ is the total number of terms in the full model, excluding the intercept and the main effect of treatment. This prior probability is given as:
\begin{equation*}
    p(m \mid \lambda_1) = \frac{e^{-\lambda_1}\lambda_1^m}{C_1 m!},
\end{equation*}
where $C_1$ is a normalization constant. We let $\boldsymbol{\omega}$ be a vector of indicators of length $p$ denoting which biomarker terms are included in a given submodel. In the absence of prior information, one can assume that all biomarker terms have equal probability of being included in a given submodel, such that $p(\boldsymbol{\omega} \mid m) = {m \choose p}^{-1}$. Similarly, we assign a prior probability to the number of knots in each spline, $k_s$:

\begin{equation*}
    p(k_s \mid \lambda_2) = \frac{e^{-\lambda_2}\lambda_2^{k_s}}{C_2 k_s!},
\end{equation*}
where $C_2$ is a normalization constant. All knot locations from a set of candidate knots of size $K_s$ have equal probability such that $p(\boldsymbol{\kappa}_s \mid k_s) = {K_s \choose k_s}^{-1}$, where $\boldsymbol{\kappa}_s$ is a vector of length $K_s$ of indicator variables denoting which knots are used for a given spline. Both $\lambda_1$ and $\lambda_2$ must be specified \textit{a priori}; larger $\lambda_1$ favors larger models and larger $\lambda_2$ favors more complex splines. The model coefficients, $\boldsymbol{\zeta}$, are assumed to have independent Normal priors, centered at $0$ with variance $\sigma_B^2$, where $\sigma_B^2$ is selected such that the prior is weakly informative.

In our simulation study, $Y_i$ is Normally distributed such that $Y_i = \mathbb{E}\left[Y_{i} \mid \mathbf{x}_i, \mathbf{z}_i, T_i\right] + \tau_i$, where $\tau_i \sim \mathcal{N}(0,\sigma_\tau^2)$. We postulate a prior distribution on $\sigma_\tau$:
\begin{equation*}
    \sigma_\tau \sim \mathcal{IG}(a_0, b_0),
\end{equation*}
where $\mathcal{IG}$ is an Inverse-Gamma distribution. 
\subsection{Computation}
\label{sec:computation}
In this section, we propose computational methods to derive the posterior distribution of all unknown parameters in Model \eqref{eq:complete}. Due to the complicated nature of the model, a closed-form analytic expression for the joint posterior distribution is unavailable and we must use a Markov Chain Monte Carlo (MCMC) sampler. To compute the posterior distribution, we  simultaneously average over the space of knot configurations for each spline term and average over the space of all combinations of  appropriate biomarker terms. Both of these elements are independently challenging as the MCMC sampler must move across spaces of varying dimension. Reversible jump (rj) MCMC, a commonly used sampler in BMA, facilitates efficient exploration of model spaces with varying dimensions \citep{chen2011bayesian}. 


The proposed rjMCMC sampling procedure proceeds as follows (see Web Appendix A for more detail). Unlike MH, which operates within a fixed parameter space and model structure, rjMCMC dynamically explores different model spaces by proposing changes in both model parameters and model dimensionality.  First, the procedure initializes parameter states. It then alternates between proposing changes in model structure using reversible jump moves (e.g., changing the knots for each spline term and adding or removing biomarker terms from the model), updating model parameters and spline coefficients within the current model using MH steps, and updating the individual-level heterogeneity parameter using a Gibbs step. In each step, a proposed state in drawn from a proposal distribution and accepted with a given probability. For reversible jump steps, the acceptance probability depends on the proposed state, the current state, and the determinant of the Jacobian matrix of the transformation to account for the change in dimensionality and ensure detailed balance. If the proposed state is accepted, the algorithm moves to the new state; otherwise, it stays in the current state. The process iterates through these steps for a large number of iterations, generating a sequence of states that characterize the posterior distribution.

More specifically, updating the set of knots for each spline term currently in the model proceeds as follows. To facilitate the description, we use the term \textit{active set} and \textit{inactive set} to refer to the set of knots or biomarker terms currently in and not in the model, respectively. The first step is a knot position change in which we propose a new location for a randomly selected knot, while keeping the number of knots constant. As this does not require a dimension change, we can use a traditional MH acceptance probability. Next, a stochastic decision is made to either propose the addition of a knot (birth step) or the removal of a knot (death step). This is a reversible jump step, in which the dimension of the proposed parameter space differs from the dimension of the current parameter space. For the birth step, a randomly-selected knot from the inactive set is proposed to be be added. This involves (1) stochastically generating a proposed model coefficient for the new knot, and (2) augmenting the proposal for other model coefficients associated with the spline term to account for this addition. The other coefficients must be augmented as changing the knot locations will change the interpretation of all coefficients in that spline term. Various methods have been proposed to deal with this challenging problem; we use an adaptation by \cite{moore2020bayesian} of methods first suggested by \cite{gamerman1997sampling}. Similarly, the death step is simply a reverse of the birth step. 

The next reversible jump step of the proposed procedure facilitates the averaging of posterior distributions of parameters across the complete space of candidate biomarkers. Here, we randomly decide between proposing to add a biomarker term (birth step) or to remove a biomarker term (death step). Rather than adding or removing one model coefficient at a time as in traditional BMA, we perform operations on the terms themselves; namely, the removal or addition of a spline term will generally correspond to a change in dimension greater than one. Further, the terms to be added or removed must be carefully selected to ensure well-formulated hierarchy is maintained. For example, if the current submodel does not contain the main effect of age, its interaction effect is not eligible to be added. It is only after the main effect is in the active set of a submodel that we can add the corresponding interaction effect. Similarly, we cannot remove a main effect term without removing its interaction effect.

For the birth step, we select an eligible term from the inactive set and propose its addition to the model. Once the eligible term is selected, we propose updates to the associated model coefficients. This involves (1) stochastically generating one or more proposed coefficients associated with the added term, and (2) proposing adjustments to the other coefficients to account for this addition. While  (2) is not necessary in general BMA, our findings suggest that such adjustments significantly enhance acceptance rates, particularly when adding a term may correspond to the addition of multiple coefficients. The death step is simply the reverse of the birth step.

\subsection{Adaptive and stopping decision rules}
\label{sec:interim_analysis}

At each interim analysis, we propose using the posterior distribution of the treatment effect averaged over the enriched population to define our decision criteria. We denote by $\mathcal{D}_\ell$ the data accumulated up to interim analysis $\ell$; we will use this information to make decisions to stop the trial early for superiority or futility and to adaptively enrich the trial population. We propose the following procedure at each interim analysis $\ell=1,\dots,L-1$:
 \begin{enumerate}
        \item \textbf{Identify the effective subspace}: Fit (or re-fit) Model \eqref{eq:complete} using the reversible jump sampling procedure specified in Section \ref{sec:computation}. The model fitting procedure is based on the data observed up to the given interim analysis point, $\mathcal{D}_\ell$. Identify the effective subspace, $\mathcal{X}_\ell^*$, using the definition provided in \eqref{eq:eff_subspace}, with $\mathcal{D}$ replaced by $\mathcal{D}_\ell$. Recalling that $\boldsymbol{\gamma}(\mathbf{\tilde{x}},\mathbf{\tilde{z}}) = \phi + \sum_{j=1}^{\tilde{J}} h_{2,j}(\tilde{x}_{ji}) + \boldsymbol{\beta}_2^T \mathbf{\tilde{z}}$ represents the subgroup-specific treatment effect, the effective subspace at the $\ell$-th interim analysis is
        \begin{equation*}
    \mathcal{X}_\ell^* = \left\{\mathbf{\tilde{x}}, \mathbf{\tilde{z}}: P\left( \boldsymbol{\gamma}(\mathbf{\tilde{x}},\mathbf{\tilde{z}}) > e_1 \mid \mathcal{D}_\ell\right)>1-\alpha\right\}.
\end{equation*}
        
Then compute the prevalence of the effective subspace, defined by the percentage of individuals in the trial who have biomarker values in $\mathcal{X}^*$. If this prevalence is less than some cutoff, $\pi$, stop the trial for futility. We then compute the enriched treatment effect as
        \begin{equation}
        \label{eq:delta_ell}
            \Delta_\ell = \int_{\mathcal{X}_\ell^*} \boldsymbol{\gamma}(\mathbf{\tilde{x}},\mathbf{\tilde{z}}) \hat{f}(\mathbf{\tilde{x}},\mathbf{\tilde{z}}) \partial \mathbf{\tilde{x}} \partial \mathbf{\tilde{z}},
        \end{equation}
        where $\hat{f}$ is the empirical probability distribution function of the tailoring variables.
        \item \textbf{Stop for efficacy}: Using the posterior predictive probability of trial success in the effective subspace, assess if the trial should be stopped for efficacy using the following criterion:
        \begin{equation}
        \label{eq:efficacy}
            P\left(\Delta_\ell>b_1\mid \mathcal{D}_\ell\right) > B_{1}.
        \end{equation}
        \item \textbf{Stop for futility}: Using the posterior predictive probability of trial failure in the effective subspace, assess if the trial should be stopped for futility using the following criterion:
          \begin{equation}
            P\left(\Delta_\ell<b_2\mid \mathcal{D}_\ell\right) > B_{2}.
        \end{equation}
\end{enumerate}

\begin{figure}[h]
    \centering
\includegraphics{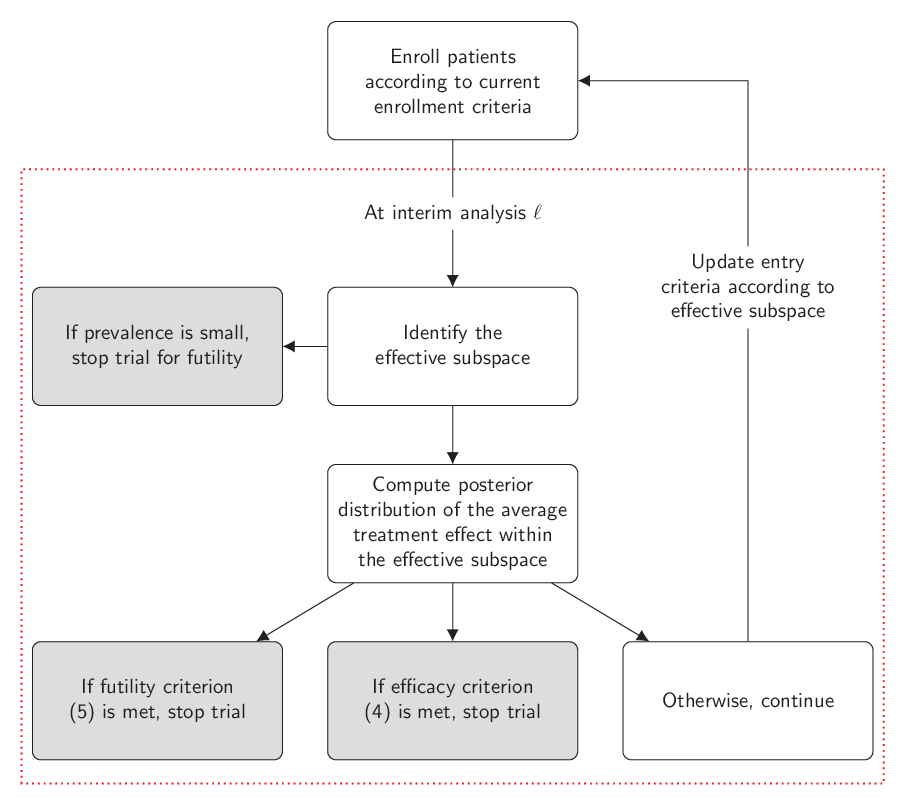}
    \caption{A visual representation of the interim decision rules}
\label{fig:interim_analysis}
\end{figure}

Figure \ref{fig:interim_analysis} summarizes the procedure. If the trial is not stopped early, at the last interim analysis, we conclude that the treatment is effective in the sensitive population using the efficacy criterion \eqref{eq:efficacy} for $\ell=L$. Otherwise, we conclude that treatment is not superior in the sensitive population. This design accounts for the possibility that the entire trial population is in the effective subspace. In this case, we assess the efficacy of treatment in the entire trial population, as in a traditional randomized clinical trial, and $\Delta_\ell$ becomes the classic average treatment effect.
We additionally incorporate adaptive enrichment after each interim analysis step. We use $\mathcal{X}_{\ell}^*$, defined in \eqref{eq:delta_ell}, to define the inclusion criteria for individuals enrolled after the $\ell$-th interim analysis. That is, after interim analysis $\ell$ only patients with variables $\mathbf{\tilde{z}},\mathbf{\tilde{x}} \in \mathcal{X}_{\ell}^*$ will be enrolled. In addition to improving the probability of enrolling patients who will benefit from treatment, we anticipate that this adaptive enrichment will improve study power. The procedure is as follows. Before the first interim analysis, there is no adaptive enrichment. After the first interim analysis, we use the accumulated data to restrict entry of patients to those expected to be in the effective subspace, updating the criteria at each interim time period.

To ensure that the effective subgroup is statistically meaningful, we may impose restrictions on the prevalence of the effective subgroup in the study. If only  a small percentage of patients benefit from the treatment, it may not be useful in practice. To prevent this, we can additionally impose that the trial should be stopped for futility if the prevalence of the effective subspace falls below some proportion $\pi$ between 0 and 1. \cite{park2022bayesian} recommend using values of $\pi$ between $0.01-0.10$; we use $\pi=0.10$ in the simulation study.

Throughout the trial, we avoid explicitly pruning the list of candidate variables as BMA will not necessarily completely drop variables which have little to no relation to the outcome. Instead, irrelevant variables will be excluded from most of the models in the sampler and have a low posterior inclusion probability. We propose to prune variables with a low posterior inclusion probability (say, $<10\%$) at the last interim analysis, or at the end of the trial. This approach of relying on fewer biomarkers for treatment recommendations not only simplifies clinical implementation but may also reduce costs, particularly if some biomarkers are costly to measure. It is entirely possible that all tailoring variables will be trimmed, which simply corresponds to a traditional, overall analysis. Finally, Model \eqref{eq:complete} can be re-fit excluding the irrelevant variables to remove unnecessary noise in preparation for clinical recommendations. 

In practice, the cutoffs ($b_1$, $b_2$, $B_1$, $B_2$) and the number and timing of interim analyses must be specified. The values of $b_1$ and $b_2$ are typically pre-specified by clinicians and correspond to clinically meaningful effect sizes. The values of $B_1$ and $B_2$ are generally tuned in design simulation studies and will depend on factors such as the number of candidate variables and their anticipated effect sizes, desired type I error rate, power, and the number of planned interim analyses. Generally, logistical considerations prevent more than three interim analyses to be conducted. A reasonable rule-of-thumb is to conduct the first interim analysis after $1/3$ or $1/2$ of the maximum sample size has been enrolled \citep{park2022bayesian}. 

\section{Application to rheumatoid arthritis}
\label{s:data_example}
We implement the proposed trial design to identify the subset of RA patients who are expected to benefit from anti-tumor necrosis factor (anti-TNF) treatment. The hypothetical trial will enroll a diverse cohort of RA patients, including those who have previously failed alternative anti-TNF therapies in the past. Patients will undergo comprehensive molecular profiling using multi-omics technologies before initiating anti-TNF treatment and at specified intervals during the treatment course. Clinical outcomes, including disease activity scores, radiographic scores, and patient-reported outcomes, will be closely monitored throughout the trial. Statistical methods, such as fitting proposed Model \eqref{eq:complete}, will be used to identify treatment-sensitive subgroups and the associated robust biomarkers. By integrating molecular data with clinical outcomes, this trial design aims to pave the way for personalized medicine approaches in managing RA, ensuring that patients receive the most effective therapies tailored to their individual biological profiles.

A meta-analysis of 41 studies investigating 99 different laboratory biomarkers revealed inconsistent associations with anti-TNF treatment response in RA patients \citep{wientjes2022prediction}. Notably, among the five biomarkers assessed in more than two studies—anti-cyclic citrullinated peptide (anti-CCP), rheumatoid factor (RF), -308 polymorphism in the TNF-$\alpha$ gene, presence of one or two shared epitope (SE) copies in the HLA-DRB1 gene, and FcGR2A polymorphism (rs1081274)—mixed results were observed. A weak association was observed for three markers studied less frequently: granulocyte-macrophage colony-stimulating factor (GM-CSF), interleukin (IL)-34 concentration, and the combined biomarker of serum IL-6 and survivin level. Building on these insights, our proposed trial design includes several clinical biomarkers, such as sex, body mass index, smoking status, alongside candidate binary and continuous molecular biomarkers. Section \ref{sec:sims} describes simulation studies to determine the operating characteristics of such a trial design. In Simulation Study I, we consider five binary biomarkers (representing, for example, sex, smoking status, RF, anti-CCP, SE copies) and one continuous marker (representing, for example, IL-34 concentration). In Simulation Study II, we consider two continuous biomarkers, such as level of GM-CSF and IL-34 concentration. 

\section{Simulation study}
\label{sec:sims}

In this section, we summarize the results of two simulation studies to assess the operating characteristics of the proposed Free Knot-Bayesian Model Averaging (FK-BMA) compared with two existing methods within the context of the RA trial. We assumed a maximum sample size of 500 with one interim analyses at 300 patients and considered candidate biomarkers which are both continuous and dichotomous. In order to define the effective subspace, we set various subsets of the variables to have a non-zero interaction with treatment; we refer to biomarkers with a non-zero interaction effect as the true tailoring variables.

We compared the proposed FK-BMA with two existing methods: Park, Liu, Thall, and Yuan [PLTY \citep{park2022bayesian}] in Simulation Study I and Liu, Kairalla, and Renfro [LKR \citep{liu2022bayesian}] in Simulation Study II. The simulations were divided into two studies, one with mixed biomarker type and another with solely continuous biomarkers, to facilitate evaluation of LKR, which can only handle continuous biomarkers. While PLTY can be used when tailoring variables are continuous, performance is poor in such settings and thus we excluded its evaluation from Study I. 

We used the fitting procedures from either method, adapted to continuous outcomes, with the interim decision rules described in Section \ref{sec:interim_analysis}. In PLTY, spike-and-slab priors are used for variable selection and each continuous biomarker is assumed to have a linear predictive relationship with outcome and a linear tailoring relationship with treatment. Additionally, well-formulated model hierarchy is maintained by first determining if the main effect should be included; if so, the interaction effect follows suit; otherwise, the interaction effect is excluded. This implies that the inclusion probability of interaction effects directly depends on their corresponding main effects' inclusion probability. Thus, the comparison between our method and PLTY emphasizes the benefits of accounting for nonlinear predictive and tailoring effects for continuous biomarkers and provides insights into the relative effectiveness of BMA and spike-and-slab priors as approaches for variable selection. To facilitate a fair comparison, we forced inclusion of the main effect of treatment in PLTY. In LKR, penalized splines are used to flexibly model the relationships between biomarkers and treatment and no variable selection is performed. The comparison between our method and LKR highlights the differences between the two spline estimation techniques, namely free-knot (FK-BMA) and penalized splines (LKR), and the benefits of variable selection in FK-BMA.

We assessed eight simulation scenarios for each study, inspired by the hypothetical RA trial, described in Table \ref{tab:sim_scenarios}. In Simulation Study I, we considered six candidate biomarkers as both predictive and tailoring variables: one continuous biomarker, $x_i$, and five binary biomarkers, $\mathbf{z}_i=(z_{1i},\dots,z_{5i})$ with prevalence 0.35, 0.50, 0.65, 0.20, and 0.35, respectively. Data were generated from 
\begin{equation}
\label{eq:gen_PLTY}
    Y_i = h_1(x_i) + \boldsymbol{\beta}_1^\top \mathbf{z}_i + \left(\phi + h_2(x_i) + \boldsymbol{\beta}_2^\top \mathbf{z}_i\right)T_i + \tau_i,
\end{equation}
for $i=1,\dots,n$ and $\tau_i \sim N(0,1)$ and $\gamma(x,\mathbf{z})=\phi + h_2(x) + \boldsymbol{\beta}_2^\top \mathbf{z}$. In Simulation Study II, we considered two continuous biomarkers $\mathbf{x}_i = (x_{1i},x_{2i})$. For $i=1,\dots,n$, the simulated individual-level data were generated from:
\begin{equation*}
    Y_i = 0.5x_{1i} +  \left(\phi + h_{2,1}(x_{1i}) + h_{2,2}(x_{2i})\right)T_i + \tau_i,
\end{equation*}
where $\tau_i \sim N(0,1)$ and $\gamma(x_1,x_2)=\phi + h_{2,1}(x_{1}) + h_{2,2}(x_{2})$. All continuous biomarkers were uniformly distributed over the interval $(0,1)$. Additional details on simulation parameters and plots are available in Web Appendix B: the true values for the predictive effects are shown in Table S1 and visual representations of scenarios with true continuous tailoring biomarkers are shown in Figures S1 and S2.

\begin{table}[]
\centering
\caption{True model parameters used for Simulation Study I and II. $\Delta$ is the true average treatment effect in the effective subspace.}
\begin{tabular}{l|ccccc}
\hline
& \multicolumn{2}{c}{Tailoring Variables} & \multicolumn{3}{c}{Effective Subspace} \\
Scenario & Type & Dimension & $\gamma(\mathbf{x},\mathbf{z})$  & Prevalence & $\Delta$ \\
\hline 
\multicolumn{6}{c}{\textit{Simulation Study I}} \\ 
    1 & - & 0 & 0 & 0 & 0 \\
    2 & - & 0 & 0.28  & 1 & 0.28\\
    3 & Binary & 1 & $z_1 - 0.3$ & 0.35 & 0.70\\
    4 & Binary & 1 & $0.7z_2 - 0.14$ & 0.50 & 0.56\\
    5 & Binary & 1 & $0.8z_3 - 0.3$ & 0.65 & 0.50\\
    6 & Binary & 2 & $0.9z_4 + 0.9z_1 - 0.2$ & 0.48 & 0.81\\
    7 & Continuous & 1 & $2.3x - 1.15$ & 0.50 & 0.58 \\
    8 & Continuous & 1 & $\text{cos}(2\pi x)$ & 0.50 & 0.64\\ 
    \hline 
    \multicolumn{6}{c}{\textit{Simulation Study II}} \\ 
    1 & - & 0 & 0 & 0 & 0 \\
    2 & - & 0 & 0.35  & 1 & 0.35\\
    3 & Continuous & 1 & $2.3x_1 - 1.15$ & 0.50 & 0.58 \\
    4 & Continuous & 1 & $\text{cos}(2\pi x_1)$ & 0.50 & 0.64\\
    5 & Continuous & 1 & $\frac{1.4\text{exp}\{25(x_1-0.5)\}}{1+\text{exp}\{25(x_1-0.5)\}}-0.6$ & 0.51 & 0.71\\
    6 & Continuous & 1 & If $x_1\le 0.5$, $\frac{2\text{exp}\{30(x_1-0.3)\}}{1+\text{exp}\{30(x_1-0.3)\}}-1$ & 0.40 & 0.77\\
    & & & If $x_1>0.5$, $\frac{2}{1+\text{exp}\{30(x_1-0.7)\}}-1$ & & \\
    7 & Continuous & 1 & If $x_1\le 0.5$, $\frac{1.5}{1+\text{exp}\{30(x_1-0.3)\}}-0.75$&  0.60 & 0.64\\
    & & & If $x_1>0.5$, $\frac{1.5\text{exp}\{30(x_1-0.7)\}}{1+\text{exp}\{30(x_1-0.7)\}}-0.75$  & & \\
    8 & Continuous & 2 & $2.3x_1  + \text{cos}(2\pi x_2) - 1.15$ & 0.50 & 0.79\\
    \hline
\end{tabular}
\label{tab:sim_scenarios}
\end{table}

For both simulation studies, we set $e_1=b_1=b_2=0$ and $\pi=0.10$ so that the trial is stopped if less than $10\%$ of patients are sensitive to treatment. We set the overall type I error rate for FK-BMA to be $0.05$ and tuned the design parameters accordingly ($B_1 = 0.975$; $B_2 = 0.8$; $\alpha=0.3$ for I and $\alpha=0.2$ for II). It is important to note that the trial parameters were specifically tuned for FK-BMA to facilitate a direct comparison of the analysis models. In practical applications, one should adjust the trial parameters according to the specific analysis model employed. We additionally tuned the MCMC parameters in each sampler for optimal performance. In Simulation Study I, we set the parameters $\lambda_1=0.1$, $\lambda_2=1$, and $\sigma_B=10$ for the FK-BMA method, tuned based on simulation. For PLTY, we tuned via simulation, as per their established notation, sparsity parameters $\mu$ and $\tau$ (Web Appendix C, Table S2). In Simulation Study II, we set $\lambda_1=0.01$, $\lambda_2=1$, and $\sigma_B=10$ for the FK-BMA approach and used 5 internal knots placed at the $(q/6)$-quantiles ($q=1,\dots,5$) for each continuous candidate biomarker in LKR method, tuned by simulation. The candidate internal knots for FK-BMA were placed at the $(q/10)$-quantiles ($q=1,\dots,9$) for each continuous candidate biomarker. We generated 1000 simulated trials in each scenario considered. Web Appendix D describes the MCMC configuration settings used in the simulation study and diagnostics to assess convergence in our proposed procedure.

As operating characteristics, we considered traditional type I error, power, generalized power to detect subgroup-specific effects, success rate in correctly identifying the appropriate tailoring variables, and accuracy of treatment recommendations using a large, external testing dataset. More specifically, we define generalized power, similar to that of \cite{chapple2018subgroup}, as the empirical probability that the following two events occur: (1) the efficacy criteria is met \eqref{eq:efficacy} in either an interim analysis step or the final analysis, and (2) the correct tailoring variables are identified. To assess accuracy of treatment recommendations, we generated an external dataset with 10,000 individuals and their biomarker information at the start of the simulation study. In each simulation iteration, we used the simulated trial results to recommend a treatment plan for individuals in the external dataset. Since the true functional relationship between biomarkers and the effect of treatment is known, we then calculated the percentage of individuals for which this treatment recommendation was optimal. 

\begin{table}[]
\centering
\caption{Results from 1000 simulated datasets in Simulation Study I for the proposed FK-BMA and the PLTY approach.}
\label{tab:PLTY_1}
\scalebox{0.8}{
\begin{tabular}{lcccccccccc}
\hline
& \multicolumn{2}{c}{Power} & \multicolumn{2}{c}{Generalized Power} & \multicolumn{2}{c}{Correct Marker} & \multicolumn{2}{c}{Accuracy} & \multicolumn{2}{c}{Size}\\
Scenario & FK-BMA & PLTY & FK-BMA & PLTY & FK-BMA & PLTY & FK-BMA & PLTY & FK-BMA & PLTY \\
\hline
&\multicolumn{10}{c}{$\mu=100$, $\tau = 0.05$} \\ 
  1 & 0.04 & 0.05 & - & - & 1.00 & 0.91 & 0.83 & 0.81 & 349.20 & 351.00 \\ 
  2 & 0.86 & 0.89 & 0.86 & 0.79 & 1.00 & 0.90 & 0.96 & 0.97 & 361.60 & 360.60 \\ 
  3 & 0.88 & 0.70 & 0.87 & 0.66 & 0.96 & 0.92 & 0.95 & 0.88 & 375.00 & 413.60 \\ 
  4 & 0.89 & 0.93 & 0.73 & 0.84 & 0.79 & 0.91 & 0.80 & 0.83 & 369.20 & 388.20 \\ 
  5 & 0.92 & 0.93 & 0.80 & 0.86 & 0.86 & 0.92 & 0.89 & 0.92 & 367.40 & 383.20 \\ 
  6 & 1.00 & 0.93 & 0.73 & 0.65 & 0.73 & 0.70 & 0.82 & 0.56 & 324.40 & 370.80 \\ 
  7 & 0.98 & 0.37 & 0.98 & 0.36 & 1.00 & 0.94 & 0.92 & 0.66 & 322.60 & 389.00 \\ 
  8 & 0.98 & 0.04 & 0.98 & 0.02 & 1.00 & 0.21 & 0.89 & 0.50 & 322.82 & 359.00 \\ 
 
& \multicolumn{10}{c}{$\mu=50$, $\tau = 0.1$} \\ 
1 & 0.03 & 0.04 & - & - & 1.00 & 0.98 & 0.80 & 0.79 & 352.80 & 351.80 \\ 
  2 & 0.89 & 0.90 & 0.89 & 0.88 & 1.00 & 0.98 & 0.97 & 0.97 & 356.80 & 359.60 \\ 
  3 & 0.86 & 0.58 & 0.86 & 0.57 & 0.96 & 0.98 & 0.95 & 0.82 & 372.00 & 419.40 \\ 
  4 & 0.88 & 0.81 & 0.72 & 0.79 & 0.78 & 0.98 & 0.80 & 0.70 & 372.80 & 411.80 \\ 
  5 & 0.92 & 0.84 & 0.80 & 0.82 & 0.86 & 0.98 & 0.90 & 0.82 & 364.00 & 407.60 \\ 
  6 & 1.00 & 0.87 & 0.70 & 0.75 & 0.71 & 0.86 & 0.82 & 0.52 & 322.20 & 382.40 \\ 
  7 & 0.98 & 0.33 & 0.98 & 0.33 & 1.00 & 0.97 & 0.92 & 0.65 & 322.82 & 393.00 \\ 
  8 & 0.98 & 0.03 & 0.98 & 0.00 & 0.99 & 0.07 & 0.88 & 0.50 & 318.02 & 358.60 \\ 
\hline
\end{tabular}
}
\end{table}

Results from Simulation Study I are presented in Table \ref{tab:PLTY_1} for two sets of PLTY sparsity parameters. The tuning parameters for FK-BMA are held constant; any observed discrepancies can be attributed to simulation variation.  In Scenario 1, both methods control type I error; in Scenario 2 (constant, positive treatment effect), they perform similarly across metrics. In scenarios with one or more binary tailoring variables (3-6), PLTY has similar or higher rates of correct marker detection, while FK-BMA has similar or higher accuracy rates. This is emphasized in Scenario 6, where the accuracy of PLTY is 0.52-0.56, compared with 0.82 for FK-BMA. Generalized power results differ in settings with a single binary tailoring variable. Specifically, in Scenario 3 with lower prevalence (0.35), FK-BMA demonstrates superior generalized power; the reverse is true in Scenarios 4-5 with higher prevalence (ranging from 0.5 to 0.65). In Scenario 6 with two binary tailoring variables, PLTY exhibits higher generalized power for the first set of sparsity parameters, while the second set favors FK-BMA. In Scenarios 7-8, there is a single continuous tailoring variable with either a linear (7) or nonlinear (8) relationship with treatment effect. With either set of sparsity parameters, the proposed FK-BMA out-performs PLTY on all metrics considered. Despite high rates of correct biomarker detection in Scenario 7, the power and accuracy of PLTY remain low. As expected, when trends are nonlinear (8), performance is poor across all metrics for PLTY as its assumptions are not met.

\begin{table}[]
\centering
\caption{Results from 1000 simulated datasets in Simulation Study I (with no predictive biomarkers, i.e., $h_1(x)=0$ and $\boldsymbol{\beta}_1=\mathbf{0}$) for the proposed FK-BMA and the PLTY approach.}
\label{tab:PLTY_2}
\scalebox{0.8}{
\begin{tabular}{lcccccccccc}
  \hline
 & \multicolumn{2}{c}{Power} & \multicolumn{2}{c}{Generalized Power} & \multicolumn{2}{c}{Correct Marker} & \multicolumn{2}{c}{Accuracy} & \multicolumn{2}{c}{Size}\\
 Scenario & FK-BMA & PLTY & FK-BMA & PLTY & FK-BMA & PLTY & FK-BMA & PLTY & FK-BMA & PLTY \\
  \hline
1 & 0.04 & 0.04 & - & - & 1.00 & 0.89 & 0.81 & 0.76 & 352.80 & 365.80 \\ 
  2 & 0.88 & 0.89 & 0.88 & 0.79 & 1.00 & 0.88 & 0.97 & 0.97 & 362.20 & 360.60 \\ 
  3 & 0.82 & 0.15 & 0.81 & 0.04 & 0.92 & 0.14 & 0.93 & 0.55 & 374.00 & 381.40 \\ 
  4 & 0.84 & 0.66 & 0.59 & 0.06 & 0.67 & 0.10 & 0.75 & 0.50 & 385.20 & 398.40 \\ 
  5 & 0.90 & 0.68 & 0.71 & 0.08 & 0.76 & 0.12 & 0.86 & 0.62 & 373.20 & 394.60 \\ 
  6 & 0.99 & 0.86 & 0.55 & 0.01 & 0.56 & 0.01 & 0.77 & 0.48 & 339.40 & 367.20 \\ 
  7 & 0.98 & 0.17 & 0.98 & 0.15 & 1.00 & 0.77 & 0.92 & 0.55 & 323.60 & 372.80 \\ 
  8 & 0.98 & 0.04 & 0.98 & 0.00 & 1.00 & 0.03 & 0.88 & 0.50 & 317.20 & 355.00 \\ 
   \hline
\end{tabular}
}
\end{table}

Recall that the comparator PLTY establishes well-formulated model hierarchy by linking the inclusion probability of interaction terms with that of the corresponding main effect terms. Specifically, if a main effect term is absent from the submodel, the interaction effect is automatically excluded. To evaluate the robustness of this approach, we conducted simulations where there are no biomarker effects observed in the control group, i.e., $h_1(x)=0$ and $\boldsymbol{\beta}_1=\mathbf{0}$ in  \eqref{eq:gen_PLTY}, using PLTY sparsity parameters $(\mu,\tau)=(100,0.05)$ (Table \ref{tab:PLTY_2}). In FK-BMA, the performance declined compared with Table \ref{tab:PLTY_1} in the case of true binary tailoring variables (Scenarios 3-6); otherwise, the results are similar. For most of the scenarios with subgroup differences (3-6, 8), the rates of detecting the correct biomarker are low, corresponding to low generalized power and low accuracy in PLTY. In Scenario 7, the rate of detecting the correct biomarker is high (0.77), while power is low (0.17). As expected, performance of PLTY remains adequate when there are no subgroup specific effects, as the set of tailoring variables (none) matches the set of predictive variables (none). Hence, ensuring a correspondence between predictive and tailoring variables is essential to achieve satisfactory performance in PLTY.

\begin{table}[]
\centering
\caption{Results from 1000 simulated datasets in Simulation Study II for the proposed FK-BMA and the LKR approach.}
\label{tab:LKR_2}
\scalebox{0.8}{
\begin{tabular}{lcccccccccc}
  \hline
 & \multicolumn{2}{c}{Power} & \multicolumn{2}{c}{Generalized Power} & \multicolumn{2}{c}{Correct Marker} & \multicolumn{2}{c}{Accuracy} & \multicolumn{2}{c}{Size}\\
 Scenario & FK-BMA & LKR & FK-BMA & LKR & FK-BMA & LKR & FK-BMA & LKR & FK-BMA & LKR \\
  \hline
1 & 0.04 & 0.24 & - & - & 1.00 & 0.00 & 0.87 & 0.86 & 334.20 & 375.83 \\ 
  2 & 0.96 & 0.99 & 0.96 & 0.00 & 1.00 & 0.00 & 0.99 & 0.79 & 329.20 & 310.28 \\ 
  3 & 0.93 & 1.00 & 0.93 & 0.00 & 0.99 & 0.00 & 0.89 & 0.91 & 325.80 & 301.84 \\ 
  4 & 0.94 & 1.00 & 0.93 & 0.00 & 0.99 & 0.00 & 0.85 & 0.89 & 327.43 & 300.60 \\ 
  5 & 0.94 & 1.00 & 0.93 & 0.00 & 0.98 & 0.00 & 0.92 & 0.94 & 334.20 & 300.62 \\ 
  6 & 0.95 & 1.00 & 0.95 & 0.00 & 0.99 & 0.00 & 0.93 & 0.94 & 310.03 & 300.41 \\ 
  7 & 0.88 & 1.00 & 0.84 & 0.00 & 0.90 & 0.00 & 0.81 & 0.88 & 353.16 & 301.21 \\ 
  8 & 0.99 & 1.00 & 0.98 & 1.00 & 0.99 & 1.00 & 0.85 & 0.89 & 308.32 & 300.00 \\ 

   \hline
\end{tabular}
}
\end{table}

Simulation Study II results, depicted in Table \ref{tab:LKR_2}, reveal LKR's notably inflated type I error (0.24) and near-maximal power in non-null scenarios.  Generalized power is zero for LKR in all but Scenario 8, as it does not perform tailoring variable selection. For FK-BMA, the type I error rate is 0.04, with lower power. Further, FK-BMA selects the correct biomarker over 90\% of the time in all scenarios considered. LKR is more sensitive to subgroup differences due to its use of a saturated, flexible model, meaning it often detects subgroup differences where they don't exist, such as in Scenario 2. Here, LKR concludes there are subgroup differences 90\% of the time (not shown), leading to a decrease in accuracy compared with FK-BMA (0.79 vs 0.99). In scenarios with subgroup-specific effects, FK-BMA demonstrates slightly lower but comparable accuracy rates to LKR, indicating a minimal loss of predictive performance from FK-BMA's model simplification.


\section{Discussion}
\label{s:discussion}
In this work, we propose Bayesian adaptive enrichment methods that effectively identify important tailoring variables from a candidate pool, flexibly model continuous biomarker effects using free knot B-splines, and reduce dependence on a single model specification by marginalizing over all possible variable combinations. Our proposed approach is the first of its kind to combine BMA with free-knot splines, leveraging the strength of both approaches. BMA integrates model uncertainty, leading to more robust inference and mitigating overfitting risks. Free knot B-splines automatically determine knot number and positions, offering a data-driven, flexible modeling approach without subjective bias. Additionally, we propose a design that incorporates adaptive enrichment and early stopping of treatment arms, potentially reducing trial costs and improving patient outcomes.

Compared with existing approaches, our proposed Bayesian adaptive enrichment design has desirable operating characteristics. Our proposed approach demonstrates comparable performance to \cite{park2022bayesian} in detecting binary tailoring variables. Notably, our method exhibits enhanced operating characteristics in scenarios involving continuous tailoring variables, including cases with nonlinear and nonmonotone effects where the assumptions of \cite{park2022bayesian} are not met, as well as in instances with simple linear effects. Additionally, our method maintains good performance in settings where a biomarker shows no association with the outcome in the control group. Thus, either our method or that of \cite{park2022bayesian} is appropriate in settings where (1) all candidate biomarkers are binary and (2) the set of true predictive biomarkers is expected to exactly match the set of true tailoring biomarkers. If one or more of these conditions is not met, our approach is more appropriate. We also compared our method with \cite{liu2022bayesian}, which uses penalized splines without variable selection. \cite{liu2022bayesian} is more sensitive to minor subgroup differences, leading to higher type I error rates and increased power. Tuning trial parameters to \cite{liu2022bayesian} would mitigate the type I error rate concern, making it suitable for settings with predetermined tailoring variables or a single candidate tailoring variable. However, our method performs well across all settings, evidenced by its comparable accuracy and high rates of correct biomarker detection.


A practical limitation of our proposed design is that it is highly computationally intensive, particularly when the number of candidate biomarkers is large. Specifically, when using binary biomarkers, computational time is reduced due to simplified model fitting. However, adding continuous markers with free-knot splines, increases computational complexity, underscoring the trade-offs between biomarker types and computational efficiency. Additionally, estimating quantities such as the empirical probability density function can be challenging within this context. Future work will explore the consequences of high dimensional variable spaces and investigate alternative sampling algorithms, such as reversible jump piece-wise deterministic processes, which may be more computationally efficient \citep{chevallier2022reversible}. 
Moreover, the effective subspace may not lead to straightforward thresholds, as we did not enforce monotonicity constraints on spline terms. To address this, methods like Bayesian additive regression trees can be employed to derive simplified cutoffs suitable for clinical applications. Additionally, we assume that the primary endpoint of the trial can be observed relatively quickly compared to the enrollment rate, ensuring that interim analyses and adaptations are well-informed by early outcomes. However, if the primary endpoint is not immediately observable, one can use early response outcomes as in \cite{park2022bayesian}. Future work will focus on adapting our methods to incorporate this proxy outcome approach. Additionally, this design should only be used in settings with a large enough sample size to detect clinically meaningful benefits. Conducting simulation studies specific to the disease context in which the trial will be conducted is crucial to determine the feasibility of such an approach.

Although our design is for a single treatment comparison, it can be extended to compare multiple treatments. Following the methods proposed by \cite{chapple2018subgroup}, one option is to introduce a latent treatment variable with strategic priors to allow treatments to be flexibly combined throughout the MCMC procedure, allowing for efficiency gains without sacrificing flexibility. Future work will formalize the model specification and assess the performance of the proposed approach in this more complex setting. 

In sum, personalized medicine offers a promising avenue to improve patient outcomes and reduce healthcare costs.  This dynamic field necessitates new ways of approaching patient recruitment and trial design, particularly in domains like RA where tailoring biomarkers and their functional forms are largely unknown but are the focus of intense investigation. To this end, we offer innovative analysis methods, computational tools, and a trial design framework, presenting an important pathway to address these evolving needs.



\section*{Acknowledgements}
Shirin Golchi is a Fonds de Recherche du Qu\'ebec, Sant\'e, Chercheuse-boursiere (Junior 1) and acknowledges support from the Natural Sciences and Engineering Council of Canada (NSERC-Discovery Grant), Canadian Statistical Sciences Institute and Fonds de Recherche du Qu\'ebec, Nature et technologies (FRQNT-NSERC NOVA). Erica E.M. Moodie is a Canada Research Chair (Tier 1) in Statistical Methods for Precision Medicine and acknowledges the support of a Chercheur de M\'erite Career Award from the Fonds de Recherche du Qu\'ebec, Sant\'e. Research reported in this publication was supported in part by the National Institute of Mental Health of the National Institutes of Health under Award Number R01 MH114873 and the McGill Interdisciplinary Initiative in Infection and Immunity.
\vspace*{-8pt}


\section*{Supplementary Materials}

Web Appendix A-D, referenced in Sections \ref{sec:computation} and \ref{sec:sims}, are available with
this paper at the Biometrics website on Wiley Online
Library. The R code to implement
the proposed methods can be found at \href{https://github.com/laramaleyeff1/Bayesian_free_knot_model_average_adaptive_design}{https://github.com/laramaleyeff1/Bayesian\_free\_knot\_model\_average\_adaptive\_design}. \vspace*{-8pt}



\bibliographystyle{biom} \bibliography{main}

\begin{thebibliography}{}

\bibitem[\protect\citeauthoryear{Adams and Brantner}{Adams and Brantner}{2006}]{adams2006estimating}
Adams, C.~P. and Brantner, V.~V. (2006).
\newblock Estimating the cost of new drug development: is it really \$802 million?
\newblock {\em Health Affairs} {\bf 25,} 420--428.

\bibitem[\protect\citeauthoryear{Asano and Hirakawa}{Asano and Hirakawa}{2020}]{asano2020bayesian}
Asano, J. and Hirakawa, A. (2020).
\newblock A {B}ayesian basket trial design accounting for uncertainties of homogeneity and heterogeneity of treatment effect among subpopulations.
\newblock {\em Pharmaceutical Statistics} {\bf 19,} 975--1000.

\bibitem[\protect\citeauthoryear{Brannath, Zuber, Branson, Bretz, Gallo, Posch, and Racine-Poon}{Brannath et~al.}{2009}]{brannath2009confirmatory}
Brannath, W., Zuber, E., Branson, M., Bretz, F., Gallo, P., Posch, M., and Racine-Poon, A. (2009).
\newblock Confirmatory adaptive designs with {B}ayesian decision tools for a targeted therapy in oncology.
\newblock {\em Statistics in Medicine} {\bf 28,} 1445--1463.

\bibitem[\protect\citeauthoryear{Chapple and Thall}{Chapple and Thall}{2018}]{chapple2018subgroup}
Chapple, A.~G. and Thall, P.~F. (2018).
\newblock Subgroup-specific dose finding in phase {I} clinical trials based on time to toxicity allowing adaptive subgroup combination.
\newblock {\em Pharmaceutical Statistics} {\bf 17,} 734--749.

\bibitem[\protect\citeauthoryear{Chen and Lee}{Chen and Lee}{2020}]{chen2020bayesian}
Chen, N. and Lee, J.~J. (2020).
\newblock {B}ayesian cluster hierarchical model for subgroup borrowing in the design and analysis of basket trials with binary endpoints.
\newblock {\em Statistical Methods in Medical Research} {\bf 29,} 2717--2732.

\bibitem[\protect\citeauthoryear{Chen, Wang, and McKeown}{Chen et~al.}{2011}]{chen2011bayesian}
Chen, X., Wang, Z.~J., and McKeown, M.~J. (2011).
\newblock A bayesian lasso via reversible-jump mcmc.
\newblock {\em Signal Processing} {\bf 91,} 1920--1932.

\bibitem[\protect\citeauthoryear{Chevallier, Fearnhead, and Sutton}{Chevallier et~al.}{2023}]{chevallier2022reversible}
Chevallier, A., Fearnhead, P., and Sutton, M. (2023).
\newblock Reversible jump {PDMP} samplers for variable selection.
\newblock {\em Journal of the American Statistical Association} {\bf 118,} 2915--2927.

\bibitem[\protect\citeauthoryear{Chu and Yuan}{Chu and Yuan}{2018}]{chu2018bayesian}
Chu, Y. and Yuan, Y. (2018).
\newblock A {B}ayesian basket trial design using a calibrated {B}ayesian hierarchical model.
\newblock {\em Clinical Trials} {\bf 15,} 149--158.

\bibitem[\protect\citeauthoryear{Freidlin, Jiang, and Simon}{Freidlin et~al.}{2010}]{freidlin2010cross}
Freidlin, B., Jiang, W., and Simon, R. (2010).
\newblock The cross-validated adaptive signature design.
\newblock {\em Clinical Cancer Research} {\bf 16,} 691--698.

\bibitem[\protect\citeauthoryear{Freidlin and Simon}{Freidlin and Simon}{2005}]{freidlin2005adaptive}
Freidlin, B. and Simon, R. (2005).
\newblock Adaptive signature design: an adaptive clinical trial design for generating and prospectively testing a gene expression signature for sensitive patients.
\newblock {\em Clinical Cancer Research} {\bf 11,} 7872--7878.

\bibitem[\protect\citeauthoryear{Gamerman}{Gamerman}{1997}]{gamerman1997sampling}
Gamerman, D. (1997).
\newblock Sampling from the posterior distribution in generalized linear mixed models.
\newblock {\em Statistics and Computing} {\bf 7,} 57--68.

\bibitem[\protect\citeauthoryear{Gu, Chen, Wei, Liu, Papadimitrakopoulou, Herbst, and Lee}{Gu et~al.}{2016}]{gu2016bayesian}
Gu, X., Chen, N., Wei, C., Liu, S., Papadimitrakopoulou, V.~A., Herbst, R.~S., and Lee, J.~J. (2016).
\newblock {B}ayesian two-stage biomarker-based adaptive design for targeted therapy development.
\newblock {\em Statistics in Biosciences} {\bf 8,} 99--128.

\bibitem[\protect\citeauthoryear{Jenkins, Stone, and Jennison}{Jenkins et~al.}{2011}]{jenkins2011adaptive}
Jenkins, M., Stone, A., and Jennison, C. (2011).
\newblock An adaptive seamless phase {II/III} design for oncology trials with subpopulation selection using correlated survival endpoints.
\newblock {\em Pharmaceutical Statistics} {\bf 10,} 347--356.

\bibitem[\protect\citeauthoryear{Jiang, Freidlin, and Simon}{Jiang et~al.}{2007}]{jiang2007biomarker}
Jiang, W., Freidlin, B., and Simon, R. (2007).
\newblock Biomarker-adaptive threshold design: a procedure for evaluating treatment with possible biomarker-defined subset effect.
\newblock {\em Journal of the National Cancer Institute} {\bf 99,} 1036--1043.

\bibitem[\protect\citeauthoryear{Jin, Riviere, Luo, and Dong}{Jin et~al.}{2020}]{jin2020bayesian}
Jin, J., Riviere, M.-K., Luo, X., and Dong, Y. (2020).
\newblock {B}ayesian methods for the analysis of early-phase oncology basket trials with information borrowing across cancer types.
\newblock {\em Statistics in Medicine} {\bf 39,} 3459--3475.

\bibitem[\protect\citeauthoryear{Karuri and Simon}{Karuri and Simon}{2012}]{karuri2012two}
Karuri, S.~W. and Simon, R. (2012).
\newblock A two-stage {B}ayesian design for co-development of new drugs and companion diagnostics.
\newblock {\em Statistics in Medicine} {\bf 31,} 901--914.

\bibitem[\protect\citeauthoryear{Liu, Kairalla, and Renfro}{Liu et~al.}{2022}]{liu2022bayesian}
Liu, Y., Kairalla, J.~A., and Renfro, L.~A. (2022).
\newblock {B}ayesian adaptive trial design for a continuous biomarker with possibly nonlinear or nonmonotone prognostic or predictive effects.
\newblock {\em Biometrics} {\bf 78,} 1441--1453.

\bibitem[\protect\citeauthoryear{Magnusson and Turnbull}{Magnusson and Turnbull}{2013}]{magnusson2013group}
Magnusson, B.~P. and Turnbull, B.~W. (2013).
\newblock Group sequential enrichment design incorporating subgroup selection.
\newblock {\em Statistics in Medicine} {\bf 32,} 2695--2714.

\bibitem[\protect\citeauthoryear{Mehta, Sch{\"a}fer, Daniel, and Irle}{Mehta et~al.}{2014}]{mehta2014biomarker}
Mehta, C., Sch{\"a}fer, H., Daniel, H., and Irle, S. (2014).
\newblock Biomarker driven population enrichment for adaptive oncology trials with time to event endpoints.
\newblock {\em Statistics in Medicine} {\bf 33,} 4515--4531.

\bibitem[\protect\citeauthoryear{Moore, MaWhinney, Carlson, and Kreidler}{Moore et~al.}{2020}]{moore2020bayesian}
Moore, C.~M., MaWhinney, S., Carlson, N.~E., and Kreidler, S. (2020).
\newblock A {B}ayesian natural cubic {B}-spline varying coefficient method for non-ignorable dropout.
\newblock {\em BMC Medical Research Methodology} {\bf 20,} 1--14.

\bibitem[\protect\citeauthoryear{Ohwada and Morita}{Ohwada and Morita}{2016}]{ohwada2016bayesian}
Ohwada, S. and Morita, S. (2016).
\newblock {B}ayesian adaptive patient enrollment restriction to identify a sensitive subpopulation using a continuous biomarker in a randomized phase 2 trial.
\newblock {\em Pharmaceutical Statistics} {\bf 15,} 420--429.

\bibitem[\protect\citeauthoryear{Park, Hsu, Siden, Thorlund, and Mills}{Park et~al.}{2020}]{park2020overview}
Park, J.~J., Hsu, G., Siden, E.~G., Thorlund, K., and Mills, E.~J. (2020).
\newblock An overview of precision oncology basket and umbrella trials for clinicians.
\newblock {\em CA: A Cancer Journal for Clinicians} {\bf 70,} 125--137.

\bibitem[\protect\citeauthoryear{Park, Liu, Thall, and Yuan}{Park et~al.}{2022}]{park2022bayesian}
Park, Y., Liu, S., Thall, P.~F., and Yuan, Y. (2022).
\newblock {B}ayesian group sequential enrichment designs based on adaptive regression of response and survival time on baseline biomarkers.
\newblock {\em Biometrics} {\bf 78,} 60--71.

\bibitem[\protect\citeauthoryear{Psioda, Xu, Jiang, Ke, Yang, and Ibrahim}{Psioda et~al.}{2021}]{psioda2021bayesian}
Psioda, M.~A., Xu, J., Jiang, Q., Ke, C., Yang, Z., and Ibrahim, J.~G. (2021).
\newblock {B}ayesian adaptive basket trial design using model averaging.
\newblock {\em Biostatistics} {\bf 22,} 19--34.

\bibitem[\protect\citeauthoryear{Renfro, Coughlin, Grothey, and Sargent}{Renfro et~al.}{2014}]{renfro2014adaptive}
Renfro, L.~A., Coughlin, C.~M., Grothey, A.~M., and Sargent, D.~J. (2014).
\newblock Adaptive randomized phase {II} design for biomarker threshold selection and independent evaluation.
\newblock {\em Chinese Clinical Oncology} {\bf 3,} 3489.

\bibitem[\protect\citeauthoryear{Robins}{Robins}{1997}]{robins1997causal}
Robins, J.~M. (1997).
\newblock Causal inference from complex longitudinal data.
\newblock In {\em Latent variable modeling and applications to causality}, pages 69--117. Springer.

\bibitem[\protect\citeauthoryear{Rosenblum, Luber, Thompson, and Hanley}{Rosenblum et~al.}{2016}]{rosenblum2016group}
Rosenblum, M., Luber, B., Thompson, R.~E., and Hanley, D. (2016).
\newblock Group sequential designs with prospectively planned rules for subpopulation enrichment.
\newblock {\em Statistics in Medicine} {\bf 35,} 3776--3791.

\bibitem[\protect\citeauthoryear{Simon}{Simon}{2017}]{simon2017critical}
Simon, R. (2017).
\newblock Critical review of umbrella, basket, and platform designs for oncology clinical trials.
\newblock {\em Clinical Pharmacology \& Therapeutics} {\bf 102,} 934--941.

\bibitem[\protect\citeauthoryear{Smolen, Aletaha, Anne~Barton, Emery, Firestein, Arthur~Kavanaugh, Solomon, Strand, and Yamamoto}{Smolen et~al.}{2018}]{smolen2018}
Smolen, J.~S., Aletaha, D., Anne~Barton, G. R.~B., Emery, P., Firestein, G.~S., Arthur~Kavanaugh, I. B.~M., Solomon, D.~H., Strand, V., and Yamamoto, K. (2018).
\newblock Rheumatoid arthritis.
\newblock {\em Nature Reviews Disease Primers} {\bf 4,} 18001.

\bibitem[\protect\citeauthoryear{Wang, O'Neill, and Hung}{Wang et~al.}{2007}]{wang2007approaches}
Wang, S.-J., O'Neill, R.~T., and Hung, H.~J. (2007).
\newblock Approaches to evaluation of treatment effect in randomized clinical trials with genomic subset.
\newblock {\em Pharmaceutical Statistics: The Journal of Applied Statistics in the Pharmaceutical Industry} {\bf 6,} 227--244.

\bibitem[\protect\citeauthoryear{Wientjes, den Broeder, Welsing, Verhoef, and van~den Bemt}{Wientjes et~al.}{2022}]{wientjes2022prediction}
Wientjes, M.~H., den Broeder, A.~A., Welsing, P.~M., Verhoef, L.~M., and van~den Bemt, B.~J. (2022).
\newblock Prediction of response to anti-tnf treatment using laboratory biomarkers in patients with rheumatoid arthritis: a systematic review.
\newblock {\em RMD open} {\bf 8,} e002570.

\end{thebibliography}


\begin{thebibliography}{}

\bibitem[\protect\citeauthoryear{Gamerman}{Gamerman}{1997}]{gamerman1997sampling}
Gamerman, D. (1997).
\newblock Sampling from the posterior distribution in generalized linear mixed models.
\newblock {\em Statistics and Computing} {\bf 7,} 57--68.

\bibitem[\protect\citeauthoryear{Geweke}{Geweke}{1991}]{RePEc:fip:fedmsr:148}
Geweke, J. (1991).
\newblock {Evaluating the accuracy of sampling-based approaches to the calculation of posterior moments}.
\newblock Staff Report 148, Federal Reserve Bank of Minneapolis.

\bibitem[\protect\citeauthoryear{Liu, Kairalla, and Renfro}{Liu et~al.}{2022}]{liu2022bayesian}
Liu, Y., Kairalla, J.~A., and Renfro, L.~A. (2022).
\newblock {B}ayesian adaptive trial design for a continuous biomarker with possibly nonlinear or nonmonotone prognostic or predictive effects.
\newblock {\em Biometrics} {\bf 78,} 1441--1453.

\bibitem[\protect\citeauthoryear{Moore, MaWhinney, Carlson, and Kreidler}{Moore et~al.}{2020}]{moore2020bayesian}
Moore, C.~M., MaWhinney, S., Carlson, N.~E., and Kreidler, S. (2020).
\newblock A {B}ayesian natural cubic {B}-spline varying coefficient method for non-ignorable dropout.
\newblock {\em BMC Medical Research Methodology} {\bf 20,} 1--14.

\bibitem[\protect\citeauthoryear{Park, Liu, Thall, and Yuan}{Park et~al.}{2022}]{park2022bayesian}
Park, Y., Liu, S., Thall, P.~F., and Yuan, Y. (2022).
\newblock {B}ayesian group sequential enrichment designs based on adaptive regression of response and survival time on baseline biomarkers.
\newblock {\em Biometrics} {\bf 78,} 60--71.

\end{thebibliography}


\label{lastpage}

\end{document}


\label{firstpage}

\maketitle

\section*{Web Appendix A: Additional Details on the Proposed rjMCMC Sampler}
\label{sec:sampler_details}

In this section, we describe the proposed reversible jump Markov Chain Monte Carlo (rjMCMC) sampler in detail (see Algorithm \ref{alg:rjmcmc} for an overview of the procedure). First, the procedure initializes parameter states. It
then alternates between proposing changes in model structure using reversible jump moves (\color{darkgray}{dark gray text}\color{black}), updating model parameters and spline coefficients within the current model
using Metropolis Hastings (MH) steps (black text), and updating the individual-level heterogeneity parameter using a Gibbs
step (black text). In each step, a proposed state in drawn from a proposal distribution and accepted
with a given probability. If the proposed state is accepted, the algorithm moves to the new state; otherwise, it stays in
the current state. The process iterates through these steps for a large number of iterations,
generating a sequence of states that characterize the posterior distribution

The following describes one iteration of the rjMCMC sampler. We assume a current state of values at a given iteration for the parameters of interest: $\boldsymbol{\kappa}_s$ and $k_s$ for $s=1,\dots,J+\tilde{J}$ (knot location and number of knots for each continuous term),  $\boldsymbol{\omega}$ (indicator vector denoting which coefficients are in the current submodel), $\boldsymbol{\zeta}$ (current value of coefficients).  We then generate states for the subsequent MCMC iteration using the following steps.

\begin{algorithm}
\caption{Overview of custom rjMCMC procedure; $R$ is the desired number of MCMC samples}\label{alg:rjmcmc}
\begin{algorithmic}[1]
\State Initialize knot configurations $\boldsymbol{\kappa}_1^{(0)}, \dots, \boldsymbol{\kappa}_{J+\tilde{J}}^{(0)}$
\State Initialize vector of included coefficients $\boldsymbol{\omega}^{(0)}$
\State Initialize model coefficients, $\boldsymbol{\zeta}^{0}$, consisting of the intercept ($\mu^{(0)}$), main effect of treatment ($\phi^{(0)}$), effects of binary variables ($\boldsymbol{\beta}^{0}$), and spline effects of continuous variables ($\boldsymbol{\theta}_1^{(0)},\dots, \boldsymbol{\theta}_{J+\tilde{J}}^{(0)}$)
\item Initialize individual-level heterogeneity $\sigma_\tau^{(0)}$ (for Normally distributed outcomes only)
\State Initialize $r=1$
\Repeat
\For{s currently in model, determined by $\boldsymbol{\omega}^{(t-1)}$}
\State Propose a knot position change and update $\boldsymbol{\kappa}_s^{(t)}$ with a usual MH step [Step 1(a)]
\State \color{darkgray}
Propose adding or removing a knot and update $\boldsymbol{\kappa}_s^{(t)}$ and $\boldsymbol{\theta}_s^{(t)}$ with a rjMCMC step [Step 1(b)]\color{black}
\State Propose updating spline basis coefficients, $\boldsymbol{\theta}_s^{(t)}$ with a MH step [Step 1(c)]
\EndFor
\State \color{darkgray} Propose adding or removing a term and update $\boldsymbol{\omega}^{(t)}$ and $\boldsymbol{\zeta}^{(t)}$ with a rjMCMC step (Step 2)\color{black}
\State Update $\mu^{(t)}$, $\phi^{(t)}$, and $\boldsymbol{\beta}^{(t)}$ with a MH step (Step 3)
\State Update $\sigma_\tau^{(t)}$ with a usual Gibbs step (Step 4)
\State Update $r=r+1$
\Until{$r=R$}
\end{algorithmic}
\end{algorithm}
\subsection*{Step 1}
For each spline term, $s$, currently in the submodel, we perform the following three-step procedure. If spline term $s$ is not currently in the submodel, Step 1 is skipped for that term.
\subsubsection*{Step 1(a)}
If $k_s>0$, a location change for one knot is proposed. First, a knot to be moved, $\kappa_{move}$, is chosen from the set of active knots (knots currently in the model), $\boldsymbol{\kappa}_s$. Since knots far away from $\kappa_{move}$ are unlikely to be accepted, a new knot location is uniformly selected from the set of inactive knots in the window, $(\kappa_{move}-w,\kappa_{move}+w)$, where $w$ is a tuning parameter than can be used to alter the acceptance rates. We denote the proposed set of knots as $\boldsymbol{\kappa}_s'$ and the proposed knot location as $\kappa'$. Since this does not involve a dimension change, we can use the traditional MH acceptance probability given by the formula:
        \begin{equation}
            \text{min}\left\{\frac{p(\mathbf{Y} \mid \boldsymbol{\kappa}_s',-)}{p(\mathbf{Y} \mid \boldsymbol{\kappa}_s,-)}\times \frac{p(\boldsymbol{\kappa}_s'\mid \lambda_2)}{p(\boldsymbol{\kappa}_s \mid \lambda_2)}\times \frac{p(\boldsymbol{\kappa}_s' \to \boldsymbol{\kappa}_s)}{p(\boldsymbol{\kappa}_s \to \boldsymbol{\kappa}_s')},1\right\},
        \end{equation}
        where $p(\mathbf{Y} \mid \boldsymbol{\kappa}_s',-)$ is the probability distribution function for the outcome of interest, $\mathbf{Y}=(Y_1,\dots,Y_n)^\top$, with $-$ referring to all other parameters. Here, all sets of knots with the same cardinality have the same prior distribution, so $p(\boldsymbol{\kappa}_s \mid \lambda_2)=p(\boldsymbol{\kappa}_s' \mid \lambda_2)$. The proposal ratio is solely determined by the quantity of vacant knots within the window for both $\kappa_{move}$ and $\kappa'$, as every knot holds an equal likelihood of being relocated, and each potential position within the window has the same probability of inclusion. The acceptance probability then simplifies to 
         \begin{equation}
            \text{min}\left\{\frac{p(\mathbf{Y} \mid \boldsymbol{\kappa}_s',-)}{p(\mathbf{Y} \mid \boldsymbol{\kappa}_s,-)}\times \frac{\text{No. of vacant in $(\kappa_{move}-w,\kappa_{move}+w)$}}{\text{No. vacant in $(\kappa'-w,\kappa'+w)$}},1\right\}.
        \end{equation}

\color{darkgray}
\subsubsection*{Step 1(b)}
\label{sec:1_b}
Next, we propose either a birth step (adding a knot) or a death step (removing a knot) with probabilities $b$ and $1-b$, respectively. 
    For the birth step, let $\boldsymbol{\kappa}_s \to \boldsymbol{\kappa}_s'$ be the proposed transition with $k_s=|\boldsymbol{\kappa}_s|=|\boldsymbol{\kappa}_s'|-1$. We first propose a knot to add from the discrete set of candidate knots not currently in the submodel, with probability $(K_s-k_s)^{-1}$, where $K_s$ is the total number of candidate knots. 

    When a new knot is added to a spline term, we must (1) recalculate the B-spline basis functions conditional on the new knot configuration, $\boldsymbol{\kappa}_s'$, (2) stochastically generate a new coefficient corresponding to the additional basis function, and (3) adjust the other spline coefficients to account for the change. We use an approach similar to \cite{moore2020bayesian} and \cite{gamerman1997sampling}, in which the proposed model coefficients are augmented by generalized linear model estimates conditional on both the current and proposed knot configuration.
    
    To this end, we first construct $\mathbf{Z}$, the complete design matrix for the saturated model. Let $B(\mathbf{x}_s,k_s,\boldsymbol{{\kappa}}_s)$ be the matrix of natural cubic B-spline basis functions evaluated at $\mathbf{x}_s$ with $k_s$ knots at locations $\boldsymbol{{\kappa}}_s$. Then, we can construct the design matrices for the main effect and the interaction effect spline terms, respectively: $\mathbf{X}_s = B(\mathbf{x}_s,k_s,\boldsymbol{{\kappa}}_s)$ for $1 \le s \le J$ and $\mathbf{X}_s = \text{diag}\{T_1,\dots,T_n\} B(\mathbf{\tilde{x}}_s,k_s,\boldsymbol{{\kappa}}_s)$ for $J < s \le J+\tilde{J}$; i.e., all rows with $T_i=0$ are set to be a vector of $0$'s. We then create a design matrix for the saturated model assuming all terms are included, relaxing this assumption in the next step. Let $\mathbf{Z}$ be such a matrix, defined as:
    \begin{equation}
        \mathbf{Z} = \begin{bmatrix}
            1 & T_1 & (\mathbf{X}_1)_{[1,]} & \dots & (\mathbf{X}_J)_{[1,]} & \mathbf{z}_1^\top & (\mathbf{X}_{J+1})_{[1,]}  & \dots &(\mathbf{X}_{J+\tilde{J}})_{[1,]} & \mathbf{\tilde{z}}_1^\top T_1 \\
           1 & T_2 & (\mathbf{X}_1)_{[2,]} & \dots & (\mathbf{X}_J)_{[2,]} & \mathbf{z}_2^\top & (\mathbf{X}_{J+1})_{[2,]} & \dots &(\mathbf{X}_{J+\tilde{J}})_{[2,]} & \mathbf{\tilde{z}}_2^\top T_2 \\
           \vdots & \vdots &\vdots &&\vdots & \vdots &\vdots &&\vdots&\vdots  \\ 
            1 & T_n & (\mathbf{X}_1)_{[n,]} & \dots & (\mathbf{X}_J)_{[n,]} & \mathbf{z}_n^\top & (\mathbf{X}_{J+1})_{[n,]} & \dots &(\mathbf{X}_{J+\tilde{J}})_{[n,]} & \mathbf{\tilde{z}}_n^\top T_n \\
        \end{bmatrix}.
    \end{equation}
    
    Next, we augment $\mathbf{Z}$ to account for which terms are currently included in the submodel. Recall that $\boldsymbol{\omega}$ is a vector of indicators with dimension $p$ denoting which terms are currently included in the submodel (excluding the intercept and main effect of treatment, which are always included). In $\boldsymbol{\omega}$, the indicators refer to which biomarker terms are in the model, while the columns of $\mathbf{Z}$ correspond to model coefficients. As spline terms will generally correspond to more than one coefficient and the design matrix includes the intercept and main effect of treatment, we need to augment $\boldsymbol{\omega}$ to account for this. To this end, we augment $\boldsymbol{\omega}$ to correspond to the terms in each column of $\mathbf{Z}$ by (1) adding indicator variables for the intercept and the main effect of treatment, which are always equal to one, and (2) padding with a vector of ones with dimension equal to the number of coefficients for each spline term. Let $\mathbf{1}_n$ denote a vector of ones of length $n$ and let $d_{s}$ refer to the number of columns in $\mathbf{X}_s$ (this will depend on both the number of internal knots, $k_s$, and the degree of the spline). The augmented indicator vector can be written as
    
    \begin{equation}
    \boldsymbol{\omega}_{coef}=\left(1,1,\underbrace{\omega_1\mathbf{1}_{d_1}^T,\dots,\omega_{J}\mathbf{1}_{d_J}^T}_{\text{Spline coefficients}},\underbrace{\omega_{J+1},\dots,\omega_{J+R}}_{\text{Binary coefficients}},\underbrace{\omega_{J+R+1}\mathbf{1}_{d_{J+1}}^T,\dots,\omega_{J+R+\tilde{J}}\mathbf{1}_{d_{J+\tilde{J}}}^T}_{\text{Spline coefficients}},\underbrace{\omega_{J+R+\tilde{J}+1},\dots \omega_p}_{\text{Binary coefficients}}\right).
    \end{equation}
    Then the design matrix for the current submodel is
    \begin{equation}
    \label{eq:design_mat}
        \tilde{\mathbf{Z}} =  \mathbf{Z}\cdot \text{diag}\{\boldsymbol{\omega}_{coef}\}.
    \end{equation} 
    In sum, this operation ensures that all columns in $\mathbf{Z}$ that correspond to coefficients not currently included in the submodel are multiplied by 0, effectively setting those columns to be a vector of zeros in $\tilde{\mathbf{Z}}$.
    
    We then use the current submodel coefficient estimates, $\boldsymbol{\zeta}$, to create an offset term that will allow us to more precisely update the current spline coefficients. The offset term, $\mathbf{\tilde{Y}}$, is simply
    \begin{equation}
        \mathbf{\tilde{Y}} = \tilde{\mathbf{Z}} \boldsymbol{\zeta}  - \mathbf{X}_s \boldsymbol{\theta}_s,
    \end{equation}
  where $\boldsymbol{\theta}_s$ are the current submodel coefficients for spline $s$. This can be interpreted as the current state of the fitted values, excluding the effects of spline $s$.

  Next, we use generalized linear model estimates based on the current and proposed spline to configuration to augment the proposed spline coefficients, $\boldsymbol{\theta}_s'$. In the linear case, we first compute $\mathbf{R} = \mathbf{{Y}} - \mathbf{\tilde{Y}}$ and then use the previous and proposed design matrices, $\mathbf{X}_s$ and $\mathbf{X}_s'$, respectively, to propose coefficient values for the spline basis functions. Let $\boldsymbol{\theta}_{s,GLM}'=(\mathbf{X}_s'^\top\mathbf{X}_s')^{-1}(\mathbf{X}_s')^\top \mathbf{R}$ and $\boldsymbol{\theta}_{s,GLM}=(\mathbf{X}_s^\top\mathbf{X}_s)^{-1}(\mathbf{X}_s)^\top \mathbf{R}$. Note that there is no need to incorporate information on which terms are currently included in the submodel from $\boldsymbol{\omega}$ as this step is only performed for currently included spline terms. In the nonlinear setting, we can use an offset term and standard iteratively reweighted least squares methods to estimate $\boldsymbol{\theta}_{s,GLM}'$ and $\boldsymbol{\theta}_{s,GLM}$. Then, the proposed coefficients, $\boldsymbol{\theta}_s'$, are calculated as follows. We generate a single random jump variable $v \sim N(0,\sigma_v^2)$, which will be used to propose a new coefficient of the added spline basis function. Let $j$ be the location in $\boldsymbol{\theta}_s$ where the new spline basis coefficient will be added. Then, $\boldsymbol{\theta}_{s,[-j]}'=\boldsymbol{\theta}_{s} + \boldsymbol{\theta}_{s,GLM,[-j]}' - \boldsymbol{\theta}_{s,GLM}$ and ${\theta}_{s,j}'=v + {\theta}_{s,GLM,j}'$, where the subscript $[-j]$ indicates that the $j$-th element has been deleted from the vector. This adjustment allows us to account for the augmented knot set and propose realistic values for the spline coefficients that improve acceptance rates.

   We then look to deriving the acceptance probability, which involves looking at the ratios of the likelihood, prior, and proposal distributions of the proposed set of knots and coefficients compared with their current counterparts. Here, the general form is
     \begin{equation}
            \text{min}\left\{\underbrace{\frac{p(\mathbf{Y} \mid \boldsymbol{\kappa}_s',\boldsymbol{\theta}_s',-)}{p(\mathbf{Y} \mid \boldsymbol{\kappa}_s, \boldsymbol{\theta}_s,-)}}_{Likelihoods}\times \underbrace{\frac{p(\boldsymbol{\theta}_s' \mid \sigma_B)}{p(\boldsymbol{\theta}_s \mid \sigma_B)}\times \frac{p(\boldsymbol{\kappa}_s' \mid \lambda_2)}{p(\boldsymbol{\kappa}_s \mid \lambda_2)}}_{Priors}\times \underbrace{\frac{1-b}{b} \times \frac{ p(\boldsymbol{\kappa}_s' \to \boldsymbol{\kappa}_s)}{ p(\boldsymbol{\kappa}_s \to \boldsymbol{\kappa}_s')} \times N(v,\sigma_v^2)^{-1}}_{Proposals},1\right\},
        \end{equation}
        where $N(v,\sigma_v^2)$ is the probability density function of a Normal distribution with mean $v$ and variance $\sigma_v^2$. Since the proposed submodel is augmented by a single additional random variable, the determinant of the Jacobian matrix of the transformation is one. First we simplify the ratio of prior distributions, starting with that of the spline coefficients, $\boldsymbol{\theta}_s$. Since there is one extra coefficient in the proposed knot configuration we can simplify terms:
        \begin{subequations}
            \begin{align}
            \frac{p(\boldsymbol{\theta}_s'\mid \sigma_B)}{p(\boldsymbol{\theta}_s \mid \sigma_B)}=(2\pi\sigma_B^2)^{-1/2}\text{exp}\left\{\frac{1}{2\sigma_B^2} (\boldsymbol{\theta}_s\boldsymbol{\theta}_s^\top - \boldsymbol{\theta}_s'\boldsymbol{\theta}_s'^\top)\right\}.
        \end{align}
        \end{subequations}
        
        Next, we simplify the ratio of priors for the number and location of knots. Recalling that $p(\boldsymbol{\kappa}_s \mid \lambda_2 )=p(\boldsymbol{\kappa}_s \mid k_s, \lambda_2) p(k_s \mid \lambda_2) = \frac{e^{-\lambda_2 \lambda_2^{k_s}}}{C k_s!} {K_s \choose k_s}^{-1}$, the ratio reduces to
        \begin{subequations}
             \begin{align}
            \frac{p(\boldsymbol{\kappa}_s' \mid \lambda_2)}{p(\boldsymbol{\kappa}_s \mid \lambda_2)} &=  \frac{e^{-\lambda_2}\lambda_2^{k_s+1}}{(k_s+1)!} \frac{k_s!}{e^{-\lambda_2}\lambda_2^{k_s}} {K_s \choose k_s + 1}^{-1} {K_s \choose k_s}\\
            &=  \frac{\lambda_2}{k_s+1} \frac{(k_s+1)!(K_s-k_s-1)!}{k_s!(K_s-k_s)!} \\
            &=  \frac{\lambda_2}{K_s-k_s}.
        \end{align}
        \end{subequations}
       
        We then look to the ratio of the proposal distributions. As shown previously, $p(\boldsymbol{\kappa}_s \to \boldsymbol{\kappa}_s')=(K_s-k_s)^{-1}$. For the reverse transition, $p(\boldsymbol{\kappa}_s' \to \boldsymbol{\kappa}_s)$, we are computing the probability that one knot out out $k_s + 1$ is removed, which is equal to $(k_s + 1)^{-1}$. Thus,
        \begin{equation}
            \frac{p(\boldsymbol{\kappa}_s' \to \boldsymbol{\kappa}_s)}{p(\boldsymbol{\kappa}_s \to \boldsymbol{\kappa}_s')} \times N(v,\sigma_v^2)^{-1} = \frac{(K_s-k_s)(2\pi\sigma_v)^{1/2}\text{exp}\left(\frac{v^2}{2\sigma_v^2}\right)}{k_s + 1}. 
        \end{equation}
        Putting this all together, the acceptance probability simplifies to
        \begin{equation}
    \text{min}\left\{\frac{p(\mathbf{Y} \mid \boldsymbol{\kappa}_s',\boldsymbol{\theta}_s', - )}{p(\mathbf{Y} \mid \boldsymbol{\kappa}_s,\boldsymbol{\theta}_s,- )}\times \frac{(1-b)\lambda_2\sigma_v}{b(k_s+1)\sigma_B}\times \text{exp}\left(\frac{v^2}{2\sigma_v^2} + \frac{1}{2\sigma_B^2}(\boldsymbol{\theta}_s\boldsymbol{\theta}_s^\top - \boldsymbol{\theta}_s'\boldsymbol{\theta}_s'^\top) \right), 1\right\},
\end{equation}

For the death step, a knot is randomly selected from $\boldsymbol{\kappa}_s$ to be removed, with probability $k_s^{-1}$. This results in a new proposed set of knots $\boldsymbol{\kappa}_s'$ with cardinality $k_s-1$. First, we calculate $\boldsymbol{\theta}_{s,GLM}'$ and $\boldsymbol{\theta}_{s,GLM}$ as in the birth step.  Then, letting $j$ be the location in $\boldsymbol{\theta}_s$ where the selected spline basis coefficient will be removed, the proposed coefficients, $\boldsymbol{\theta}_s'$, are calculated as follows: $\boldsymbol{\theta}_{s}'=\boldsymbol{\theta}_{s,[-j]} + \boldsymbol{\theta}_{s,GLM}' - \boldsymbol{\theta}_{s,GLM,[-j]}$, where the subscript $[-j]$ indicates that the $j$-th element has been deleted from the vector. The general form of the acceptance probability for the death step is
     \begin{equation}
            \text{min}\left\{\underbrace{\frac{p(\mathbf{Y} \mid \boldsymbol{\kappa}_s',\boldsymbol{\theta}_s'-)}{p(\mathbf{Y} \mid \boldsymbol{\kappa}_s, \boldsymbol{\theta}_s-)}}_{Likelihoods}\times \underbrace{\frac{p(\boldsymbol{\theta}_s' \mid \sigma_B)}{p(\boldsymbol{\theta}_s \mid \sigma_B)}\times \frac{p(\boldsymbol{\kappa}_s' \mid \lambda_2)}{p(\boldsymbol{\kappa}_s \mid \lambda_2)}}_{Priors}\times \underbrace{\frac{b}{1-b} \times \frac{ p(\boldsymbol{\kappa}_s' \to \boldsymbol{\kappa}_s)}{ p(\boldsymbol{\kappa}_s \to \boldsymbol{\kappa}_s')} \times N(v,\sigma_v^2)}_{Proposals},1\right\},
        \end{equation}
where $v \sim N(0,\sigma_v^2)$ is a random jump variable, generated to account for the reverse transition. Similarly to the birth step, we can simplify terms. Since there is one extra coefficient in the current knot configuration, we can reduce the prior ratios for the model coefficients:
\begin{equation}
    \frac{p(\boldsymbol{\theta}_s' \mid \sigma_B)}{p(\boldsymbol{\theta}_s \mid \sigma_B)}=(2\pi\sigma_B^2)^{1/2}\text{exp}\left\{\frac{1}{2\sigma_B^2} (\boldsymbol{\theta}_s\boldsymbol{\theta}_s^\top - \boldsymbol{\theta}_s'\boldsymbol{\theta}_s'^\top)\right\}.
\end{equation}
We can then simplify the prior ratios of the knot configurations:
\begin{subequations}
\begin{align}
    \frac{p(\boldsymbol{\kappa}_s' \mid \lambda_2)}{p(\boldsymbol{\kappa}_s\mid \lambda_2)} &=  \frac{e^{-\lambda_2}\lambda_2^{k_s-1}}{(k_s-1)!} \frac{k_s!}{e^{-\lambda_2}\lambda_2^{k_s}} {K_s \choose k_s - 1}^{-1} {K_s \choose k_s}\\
    &=  \frac{k_s}{\lambda_2} \frac{(k_s-1)!(K_s-k_s+1)!}{k_s!(K_s-k_s)!} \\
    &=  \frac{K_s-k_s+1}{\lambda_2}.
\end{align}
\end{subequations}
As shown previously, $p(\boldsymbol{\kappa}_s \to \boldsymbol{\kappa}_s')=k_s^{-1}$. For the reverse transition, $p(\boldsymbol{\kappa}_s' \to \boldsymbol{\kappa}_s)$, we are computing the probability that one knot out of $K_s-(k_s-1)$ is added, which is equal to $(K_s-k_s+1)^{-1}$. So:
 \begin{equation}
    \frac{p(\boldsymbol{\kappa}_s' \to \boldsymbol{\kappa}_s)}{p(\boldsymbol{\kappa}_s \to \boldsymbol{\kappa}_s')} = \frac{k_s}{K_s - k_s + 1}.
\end{equation}
Putting this all together, the acceptance probability becomes
\begin{equation}
    \text{min}\left\{\frac{p(\mathbf{Y} \mid \boldsymbol{\kappa}_s',\boldsymbol{\theta}_s', - )}{p(\mathbf{Y} \mid \boldsymbol{\kappa}_s,\boldsymbol{\theta}_s,- )}\times \frac{b k_s \sigma_B}{(1-b)\lambda_2 \sigma_v}\times \text{exp}\left(-\frac{v^2}{2\sigma_v^2} + \frac{1}{2\sigma_B^2}(\boldsymbol{\theta}_s\boldsymbol{\theta}_s^\top - \boldsymbol{\theta}_s'\boldsymbol{\theta}_s'^\top) \right), 1\right\}.
\end{equation}
\color{black}
\subsubsection*{Step 1(c)}
Finally, we update the spline coefficients via a traditional MH step using a random walk centered at the previous coefficient value: $\boldsymbol{\theta}_s' = \boldsymbol{\theta}_s + \boldsymbol{\epsilon}$, where $\boldsymbol{\epsilon}$ are independent and Normally distributed with mean $0$ and variance $\sigma_\epsilon^2$ with equal dimension to $\boldsymbol{\theta}_s$. The acceptance probability is then
\begin{equation}
     \text{min}\left\{\frac{p(\mathbf{Y} \mid \boldsymbol{\theta}_s', - )}{p(\mathbf{Y} \mid \boldsymbol{\theta}_s,- )}\times \frac{p(\boldsymbol{\theta}_s' \mid \sigma_B )}{p(\boldsymbol{\theta}_s \mid \sigma_B)}, 1\right\}.
\end{equation}

\color{darkgray}
\subsection*{Step 2} 
Next, we proceed with the Bayesian model averaging (BMA) step of the proposed procedure, facilitating the averaging of posterior distributions of parameters across the complete space of candidate models. Here,
we propose adding a biomarker term (birth step) or removing a biomarker term (death step), with probabilities $c$ and $1-c$, respectively. Rather than adding or removing one model coefficient at a time as in traditional BMA, we perform operations on the terms themselves; namely, the removal or addition of a spline term will generally correspond to a change in dimension greater than one. Further, the terms to be added or removed must be carefully selected to ensure well-formulated hierarchy is maintained. For example, if the current submodel does not contain the main effect of age, its interaction effect is not eligible to be added. It is only after the main effect is in the active set of a submodel that we can add the corresponding interaction effect. Similarly, we cannot remove a main effect term without removing its interaction effect.

For the birth step, let $\boldsymbol{\omega} \to \boldsymbol{\omega}'$ be the proposed transition, where $|\boldsymbol{\omega}|=m$, $|\boldsymbol{\omega}'|=m+1$. We select an eligible term from the inactive set and propose its addition to the model. Once the eligible term is selected, we propose updates to the associated model coefficients. This involves (1) stochastically generating one or more proposed coefficients associated with the added term, and (2) proposing adjustments to the other coefficients to account for this addition. Although (2) is not necessary in BMA, we found that the sampler performance was greatly improved when we  augmented coefficient proposals using an adaptation of previous methods \citep{moore2020bayesian,gamerman1997sampling}.  To this end, in the linear setting, let $\boldsymbol{\zeta}_{GLM}'=(\mathbf{\tilde{Z}}'^\top\mathbf{Z}')^{-1}(\mathbf{\tilde{Z}}')^\top \mathbf{Y}$ and $\boldsymbol{\zeta}_{GLM}=(\mathbf{\tilde{Z}}^\top\mathbf{\tilde{Z}})^{-1}(\mathbf{\tilde{Z}})^\top \mathbf{Y}$, with the construction of $\mathbf{\tilde{Z}'}$ and $\mathbf{\tilde{Z}}$ described in \eqref{eq:design_mat} of Step 1(b). In the nonlinear setting, we can use standard iteratively reweighted least squares methods to estimate $\boldsymbol{\zeta}_{GLM}'$ and $\boldsymbol{\zeta}_{GLM}$ by fitting generalized linear models relating $\mathbf{Y}$ to $\mathbf{\tilde{Z}}'$ and  $\mathbf{\tilde{Z}}$, respectively.

We then propose new values for currently included model coefficients. First, we generate jumps for the new coefficient(s): $u_i \sim N(0,\sigma_u^2)$, for $i=1,\dots,d$ where $d$ is the number of coefficients associated with the chosen term. For the coefficients that are in both the current model and the proposed model we set $\boldsymbol{\zeta}'=\boldsymbol{\zeta} + \boldsymbol{\zeta}_{GLM}' - \boldsymbol{\zeta}_{GLM}$ and for the set of coefficients that are proposed to be added we let $\boldsymbol{\zeta}'=\mathbf{u} + \boldsymbol{\theta}_{GLM}'$.  With the proposed coefficients and included terms, we can then construct the acceptance ratio:
     \begin{equation}
            \text{min}\left\{\underbrace{\frac{p(\mathbf{Y} \mid \boldsymbol{\omega}',\boldsymbol{\zeta}',-)}{p(\mathbf{Y} \mid \boldsymbol{\omega},\boldsymbol{\zeta}',-)}}_{Likelihoods}\times \underbrace{\frac{p(\boldsymbol{\zeta}' \mid \sigma_B)}{p(\boldsymbol{\zeta} \mid \sigma_B)}\times \frac{p(\boldsymbol{\omega}' \mid \lambda_1)}{p(\boldsymbol{\omega} \mid \lambda_1)}}_{Priors}\times \underbrace{\frac{1-c}{c} \times \frac{ p(\boldsymbol{\omega}' \to \boldsymbol{\omega})}{ p(\boldsymbol{\omega} \to \boldsymbol{\omega}')} \times \prod_{i=1}^d N(u_i,\sigma_u^2)^{-1}}_{Proposals},1\right\},
        \end{equation}
where $N(u,\sigma_u^2)$ is the probability density function of a Normal distribution with mean $u$ and variance $\sigma_u^2$ and $p(\boldsymbol{\omega} \to \boldsymbol{\omega}')$ and $p(\boldsymbol{\omega}' \to \boldsymbol{\omega})$ are the respective inverses of the number of variables which are eligible to be added or removed for the given transition while maintaining model hierarchy. Since the model is augmented by a set of additional random variables, the determinant of the Jacobian matrix of the transformation is one. We then simplify the ratio of priors for which coefficients are included in the model. Recalling that $p(\boldsymbol{\omega} \mid \lambda_1 )=p(\boldsymbol{\omega} \mid m, \lambda_1) p(m \mid \lambda_1) = \frac{e^{-\lambda_1 \lambda_1^{k_s}}}{C m!} {p \choose m}^{-1}$, 
\begin{subequations}
\begin{align}
        \frac{p(\boldsymbol{\omega}'\mid \lambda_1)}{p(\boldsymbol{\omega}\mid \lambda_1)} &= \frac{\lambda_1^{m+1}}{(m+1)!} \frac{m!}{\lambda_1^m} \frac{(m+1)!(p-m-1)!}{m!(p-m)!}\\
        &= \frac{\lambda_1}{p-m}.
\end{align}
\end{subequations}
Thus, the acceptance probability for a birth transition $\boldsymbol{\omega} \to \boldsymbol{\omega}'$, $\boldsymbol{\zeta} \to \boldsymbol{\zeta}'$ simplifies to

\begin{equation}
    \text{min}\left\{\frac{p(\mathbf{Y} \mid \boldsymbol{\omega}',\boldsymbol{\zeta}',- )}{p(\mathbf{Y} \mid \boldsymbol{\omega},\boldsymbol{\zeta},- )} \times \frac{p(\boldsymbol{\zeta}' \mid \sigma_B )}{p(\boldsymbol{\zeta} \mid \sigma_B)} \times \frac{(1-c)\lambda_1}{c(p - m)} \times \frac{p(\boldsymbol{\omega}' \to \boldsymbol{\omega})}{p(\boldsymbol{\omega} \to \boldsymbol{\omega}')} \times \prod_{i=1}^d N(u_i,\sigma_u^2)^{-1}  , 1\right\}.
\end{equation}

For the death step, we propose the transition $\boldsymbol{\omega} \to \boldsymbol{\omega}'$, where $|\boldsymbol{\omega}|=m$, $|\boldsymbol{\omega}'|=m-1$. Here, one variable out of the eligible inactive set is selected to be removed. Similarly to the birth step, we (1) stochastically generate a random jump variable to account for the reverse transition and (2) propose adjustments to account for the coefficient removal. For the reverse transition, we generate $u_i \sim N(0,\sigma_u^2)$, for $i=1,\dots,d$ where $d$ is the number of coefficients associated with the chosen term. For the coefficient adjustment, we compute $\boldsymbol{\zeta}_{GLM}'$ and $\boldsymbol{\zeta}_{GLM}$ as described in the birth step. We then augment the proposed coefficients. For the coefficients that are in both the current model and the proposed model we let: $\boldsymbol{\zeta}'=\boldsymbol{\zeta} + \boldsymbol{\zeta}_{GLM}' - \boldsymbol{\zeta}_{GLM}$. The coefficients to be removed from the proposed submodel are set to zero. With the proposed coefficients, we can then construct the acceptance ratio:
     \begin{equation}
            \text{min}\left\{\underbrace{\frac{p(\mathbf{Y} \mid \boldsymbol{\omega}',\boldsymbol{\zeta}',-)}{p(\mathbf{Y} \mid \boldsymbol{\omega},\boldsymbol{\zeta}',-)}}_{Likelihoods}\times \underbrace{\frac{p(\boldsymbol{\zeta}' \mid \sigma_B)}{p(\boldsymbol{\zeta} \mid \sigma_B)}\times \frac{p(\boldsymbol{\omega}' \mid \lambda_1)}{p(\boldsymbol{\omega} \mid \lambda_1)}}_{Priors}\times \underbrace{\frac{c}{1-c} \times \frac{ p(\boldsymbol{\omega}' \to \boldsymbol{\omega})}{ p(\boldsymbol{\omega} \to \boldsymbol{\omega}')} \times \prod_{i=1}^d N(u_i,\sigma_u^2)}_{Proposals},1\right\},
        \end{equation}
where $p(\boldsymbol{\omega} \to \boldsymbol{\omega}')$ and $p(\boldsymbol{\omega}' \to \boldsymbol{\omega})$ are the respective inverses of the number of variables which are eligible to be removed or added for the given transition while maintaining model hierarchy. We can simplify the ratio of the prior distributions:
\begin{subequations}
    \begin{align}
        \frac{p(\boldsymbol{\omega}'\mid m - 1)p(m - 1 \mid \lambda_1)}{p(\boldsymbol{\omega}\mid m)p(m \mid \lambda_1)} &= \frac{\lambda_1^{m-1}}{(m-1)!} \frac{m!}{\lambda_1^m} \frac{(m-1)!(p-m+1)!}{m!(p-m)!}\\
        &= \frac{(p-m+1)}{\lambda_1}.
        \end{align}
\end{subequations}
The simplified acceptance probability for the death step is
\begin{equation}
    \text{min}\left\{\frac{p(\mathbf{Y} \mid \boldsymbol{\omega}',\boldsymbol{\zeta}',- )}{p(\mathbf{Y} \mid \boldsymbol{\omega},\boldsymbol{\zeta},- )} \times \frac{p(\boldsymbol{\zeta}'\mid \sigma_B)}{p(\boldsymbol{\zeta}\mid \sigma_B)} \times \frac{c(p - m+1)}{(1-c)\lambda_1} \times \frac{p(\boldsymbol{\omega}' \to \boldsymbol{\omega})}{p(\boldsymbol{\omega} \to \boldsymbol{\omega}')} \times  \prod_{i=1}^{d}N(u_i,\sigma_u^2), 1\right\}.
\end{equation}
\color{black}
\subsection*{Step 3} 
We then update the remaining coefficients, which consist of the intercept, the main effect of treatment, and the main and interaction effects of the included binary variables, using a usual MH step. Letting $\boldsymbol{\beta} = (\boldsymbol{\beta}_1^\top,\boldsymbol{\beta}_2^\top)^\top$, we recommend using a random walk centered at the previous coefficient value: $(\mu', \phi', \boldsymbol{\beta}'^\top)^\top = (\mu, \phi, \boldsymbol{\beta}^\top)^\top  + \boldsymbol{\epsilon}$, where $\boldsymbol{\epsilon}$ are independent and Normally distributed with mean $0$ and variance $\sigma_\epsilon^2$. The acceptance probability is then,
\begin{equation}
     \text{min}\left\{\frac{p(\mathbf{Y} \mid \mu', \phi', \boldsymbol{\beta}', - )}{p(\mathbf{Y} \mid \mu, \phi, \boldsymbol{\beta},- )}\times \frac{p(\mu', \phi', \boldsymbol{\beta}' \mid \sigma_B )}{p(\mu, \phi, \boldsymbol{\beta} \mid \sigma_B)}, 1\right\}.
\end{equation}
This can also be done using separate steps for each coefficient.
\subsection*{Step 4} 
For Normally-distributed outcomes, we update the individual-level heterogeneity term, $\sigma_\tau$, using a Gibbs step:
\begin{equation}
    \sigma_\tau \sim \mathcal{IG}\left(\frac{n}{2} + a_0, \sum_{i=1}^n \frac{(Y_i - \mathbf{\tilde{Z}}_i\boldsymbol{\zeta})^2}{2} + b_0\right),
\end{equation}
where $n$ is the sample size, $\mathbf{\tilde{Z}}_i$ is the $i$-th row of the current design matrix, $\mathbf{\tilde{Z}}$, defined in \eqref{eq:design_mat} of Step 1(b), and $\boldsymbol{\zeta}$ are the current values of the model coefficients.

\section*{Web Appendix B: Simulation Set-up}

In this section, we describe additional details of the data generation processes in Simulation Study I and II. Recall that in Simulation Study I, we considered six candidate biomarkers as both predictive and tailoring variables: one continuous biomarker, $x_i$, and five binary biomarkers, $\mathbf{z}_i=(z_{1i},\dots,z_{5i})$ with prevalence 0.35, 0.50, 0.65, 0.20, and 0.35, respectively. Data were generated from 
\begin{equation}
    Y_i = h_1(x_i) + \boldsymbol{\beta}_1^\top \mathbf{z}_i + \gamma(x_i,\mathbf{z}_i)T_i + \tau_i,
\end{equation}
for $i=1,\dots,n$ and $\tau_i \sim N(0,1)$. Table \ref{tab:predictive_effects} describes the true model parameters ($h_1(x)$ and $\boldsymbol{\beta}_1$) used for the predictive effects in Simulation Study I. The predictive effects that are non-zero correspond exactly to the non-zero tailoring effects; that is, the set of true predictive biomarkers and true tailoring biomarkers are identical. The results of relaxing this assumption are shown in Table 3.

\begin{table}[ht]
\centering
  \caption{True model parameters used for the predictive effects in Simulation Study I (Table 2). In all scenarios, $\beta_{15}=0$.}
    \begin{tabular}{c|cccccccc}
    \hline
    & \multicolumn{8}{c}{Scenarios} \\
    Parameter & 1 & 2 & 3 & 4 & 5 & 6 & 7 & 8 \\
    \hline
    $\beta_{11}$ & 0 & 0 & 0.5 & 0 & 0 & 0.3 & 0 & 0  \\
        $\beta_{12}$ &  0 & 0 & 0 & 0.5 & 0 & 0 & 0 & 0  \\
        $\beta_{13}$ &  0 & 0 & 0 & 0 & 0.5 & 0 & 0 & 0  \\
        $\beta_{14}$ &  0 & 0 & 0 & 0 & 0 & 0.5 & 0 & 0  \\
        $h_1(x)$ & 0 & 0 & 0 & 0 & 0  & $0$ & $0.3x$ & $0.3x$\\
        \hline
    \end{tabular}
    \label{tab:predictive_effects}
\end{table}

\begin{figure}
    \centering
    \includegraphics[width=\textwidth]{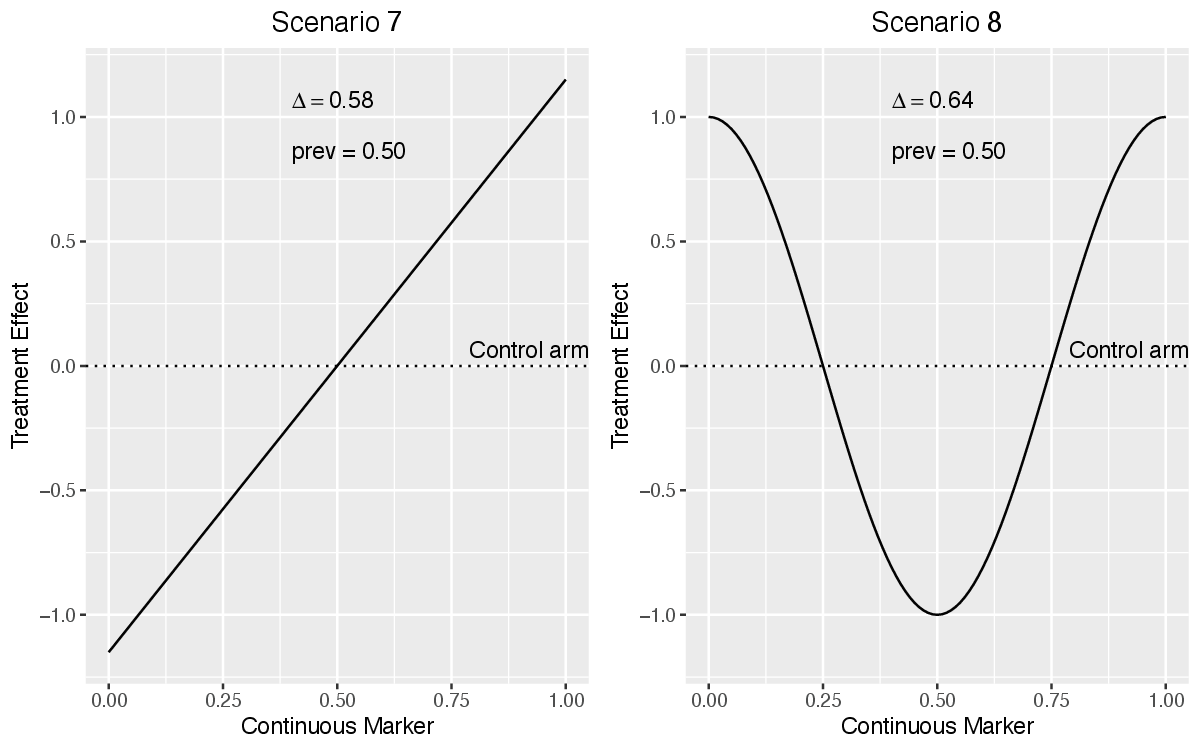}
    \caption{True effects of the treatment vs. control arms on the outcome as a function of the continuous biomarker in Simulation Study I. $\Delta$ is the true average treatment effect in the effective subspace and prev refers to the true prevalence of the effective subspace.}
    \label{fig:park_visual}
\end{figure}

Figure \ref{fig:park_visual} shows the relationship between the continuous biomarker and treatment effectiveness in Study I for the scenarios (7-8) in which there is a continuous tailoring variable. In Scenario 7, the continuous biomarker has a linear relationship with treatment effectiveness and the assumptions of PLTY are met. In Scenario 8, the continuous biomarker has a nonlinear and nonmonotone relationship with treatment effectiveness, violating the assumptions of PLTY. Scenarios 1-6 are described in Table 1.

\begin{figure}
    \centering
    \includegraphics[width=\textwidth]{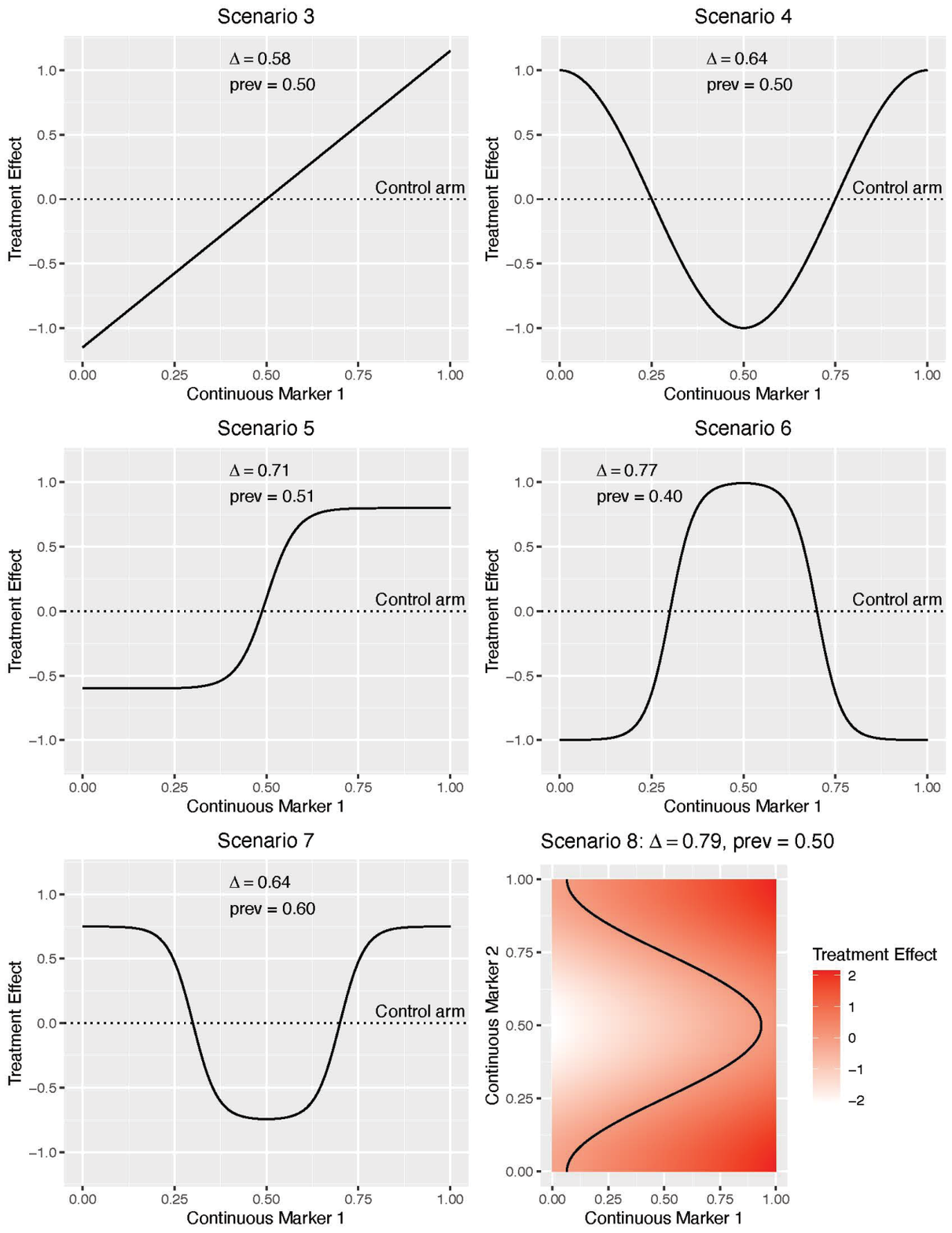}
     \caption{True effects of the treatment vs. control arms on outcome as a function of the continuous biomarkers in Simulation Study II. $\Delta$ is the true average treatment effect in the effective subspace and prev refers to the true prevalence of the effective subspace. In Scenario 8, treatment effects of 0 are denoted with a black line.}
    \label{fig:liu_visual}
\end{figure}

Figure \ref{fig:liu_visual} shows the relationship between the continuous biomarkers and treatment effectiveness for the scenarios (3-8) in Study II in which there are one or more true continuous tailoring variables. In Scenario 3, the biomarker has a linear relationship with treatment effectiveness. In Scenarios 4, 6, and 7, the biomarker has a nonlinear and nonmonotone relationship with treatment effectiveness. In Scenario 5, the biomarker has a nonlinear and monotone relationship with treatment effectiveness. In Scenario 8, both biomarkers are related to treatment effectiveness: biomarker 1 has a linear relationship and biomarker 2 has a nonlinear and nonmonotone relationship. Scenarios 1-2 are described in Table 1.

\section*{Web Appendix C: Tuning Sparsity Parameters}
Table \ref{tab:tune_park} describes the results of a simulation study to tune the sparsity parameters for PLTY. Candidate values from \cite{park2022bayesian} (Web Appendix I, Web Table 11) were used. In the final analysis, we chose to present results from $(\mu,\tau)=(100,0.05)$ and $(50, 0.10)$, selected by maximizing both generalized power and rates of detecting the correct biomarker. 
\begin{table}[ht]
\centering
\caption{Results from 1000 simulated datasets for the proposed FK-BMA and the PLTY approach with varying sparsity parameters. Data were generated under Scenario 4 of Simulation Study I.}
\label{tab:tune_park}
\scalebox{0.9}{
\begin{tabular}{llcccccccc}
  \hline
 & & \multicolumn{2}{c}{Power} & \multicolumn{2}{c}{Generalized Power} & \multicolumn{2}{c}{Correct Marker} & \multicolumn{2}{c}{Accuracy} \\
$\mu$ & $\tau$& FK-BMA & PLTY & FK-BMA & PLTY & FK-BMA & PLTY & FK-BMA & PLTY  \\
  \hline
10 & 0.10 & 0.89 & 0.91 & 0.72 & 0.59 & 0.78 & 0.66 & 0.81 & 0.88 \\ 
  50 & 0.10 & 0.88 & 0.81 & 0.72 & 0.79 & 0.78 & 0.98 & 0.80 & 0.70 \\ 
  100 & 0.05 & 0.89 & 0.93 & 0.73 & 0.84 & 0.79 & 0.91 & 0.80 & 0.83 \\ 
  100 & 0.10 & 0.88 & 0.74 & 0.74 & 0.65 & 0.79 & 0.88 & 0.81 & 0.62 \\ 
  100 & 0.50 & 0.90 & 0.55 & 0.75 & 0.00 & 0.80 & 0.00 & 0.81 & 0.50 \\ 
  100 & 1.00 & 0.89 & 0.52 & 0.75 & 0.00 & 0.81 & 0.00 & 0.81 & 0.50 \\ 
   \hline
\end{tabular}
}
\end{table}

\section*{Web Appendix D: MCMC Configuration Settings and Diagnostics}

In this section, we describe the MCMC configuration settings used in the simulation study, such as burn-in, number of iterations, and thinning factor, and suggest diagnostics which can be used to assess model convergence for our rjMCMC procedure. In the proposed rjMCMC procedure, we used a burn-in of 30,000 samples, a thinning interval of 10, one chain, and 2,000 desired posterior samples. These settings were also applied to the MCMC procedure of LKR. For the PLTY MCMC procedure, we reduced the burn-in to 10,000 samples, used one chain, set the desired number of posterior samples to 5,000, and omitted thinning. In our procedure, we initialized the coefficient and residual standard error values to be the corresponding ordinary least squares estimates, with all biomarker terms included and zero internal knots for all spline terms.

When running a rjMCMC sampler, tracking individual parameters across the sequence is difficult as they can disappear and reappear at any stage. This complicates statistical analysis and the assessment of convergence. One solution is to apply conventional MCMC diagnostics to parameters that are included in all sub-models throughout the sequence. In our model, such possibilities include the main effect of treatment, which is included in all sub-models, and the tailored treatment effect for each individual. Although the individual treatment effects may depend on different biomarkers throughout the rjMCMC sequence, the parameter itself is well-defined throughout and thus its trajectory can be tracked.

To assess convergence in the simulation study, we used a diagnostic proposed by \cite{RePEc:fip:fedmsr:148}, based on a test for equality of the means of the first and last part of a Markov chain (we used the first 25\% and the last 25\%). Large differences indicate lack of convergence. In our study, we calculated the absolute Geweke diagnostic for each individual treatment effect, and if the maximum value was below four, a threshold used in \cite{liu2022bayesian}, we concluded the rjMCMC converged.

\begin{figure}
    \centering
\includegraphics[width=\textwidth]{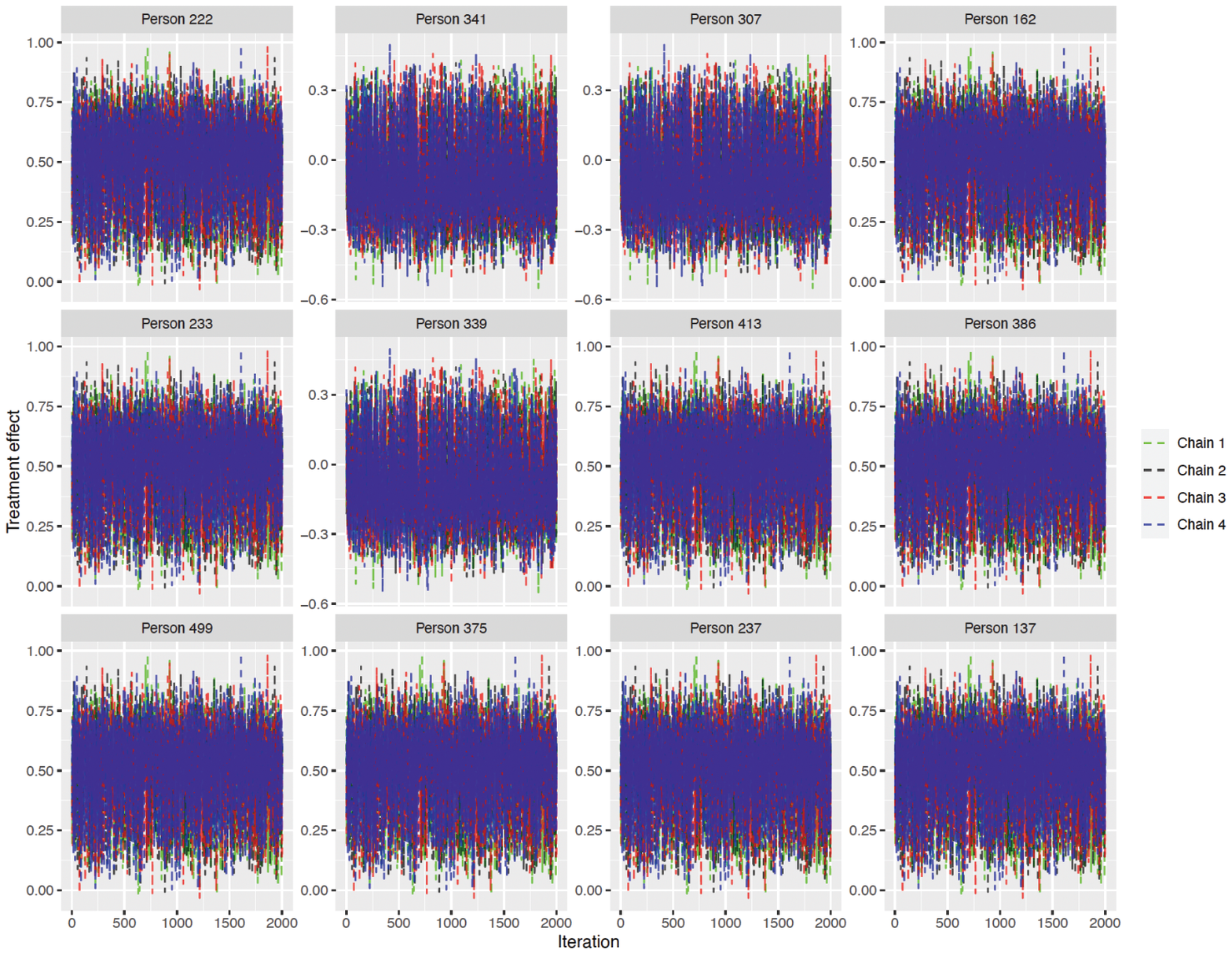}
    \caption{Trace plots for a random sample of individual treatment effects from the proposed rjMCMC procedure applied to a simulated dataset.}
    \label{fig:traceplots}
\end{figure}

\begin{figure}
    \centering
\includegraphics[width=\textwidth]{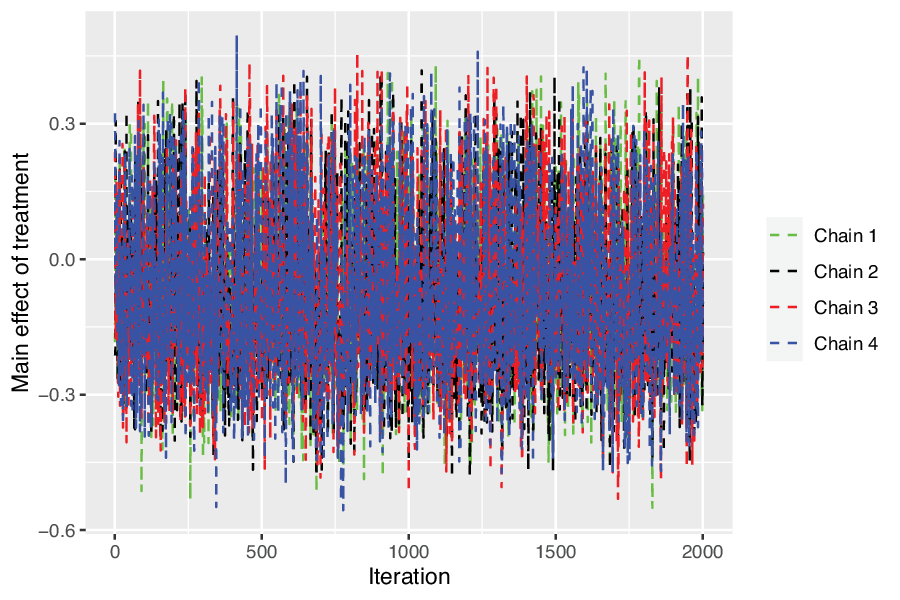}
    \caption{Trace plot for the main effect of treatment, $\phi$, from the proposed rjMCMC procedure applied to a simulated dataset.}
    \label{fig:traceplots_trt}
\end{figure}

In practice, there are several additional MCMC diagnostics which can be used to assess convergence. These include the Gelman-Rubin diagnostic ($\hat{R}$), trace plots, autocorrelation plots, and effective sample size, among others. To illustrate, we applied our proposed rjMCMC procedure to fit Model (1) using a simulated dataset with 500 individuals. The dataset was generated based on Scenario 4 of Simulation Study I, in which there is one true binary tailoring variable with a prevalence of 0.5. We ran four chains using the same MCMC configuration settings as in the simulation study. To assess convergence, we computed $\hat{R}$, Geweke diagnostics, and trace plots for the 500 individual treatment effects and the main effect of treatment, $\phi$. All $\hat{R}$ values for the individual treatment effects and $\phi$ were found to be 1.00. Geweke diagnostics ranged from -1.60 to 0.63 for both sets of effects. For trace plots and other graphical diagnostics, we recommend examining a random sample of study participants due to feasibility constraints in plotting each individual's trajectory. Figures \ref{fig:traceplots} and \ref{fig:traceplots_trt} display the trace plots for a random sample of individual treatment effects and for the main treatment effect, respectively. Standard R packages, such as \textit{coda}, can be used to generate these graphical diagnostics. Together, these diagnostics confirm the convergence of the proposed rjMCMC procedure for the simulated dataset.

\bibliographystyle{biom}
\bibliography{supp}